\documentclass[journal]{IEEEtran}
\usepackage[lined,boxed,commentsnumbered]{algorithm2e}
\usepackage[cmex10]{amsmath}
\usepackage{mathrsfs}
\usepackage[dvips]{color, graphicx}
\usepackage[lined,boxed,commentsnumbered]{algorithm2e}
\usepackage{cite}
\usepackage{graphicx}
\usepackage{subfigure}
\usepackage{url}
\usepackage{stfloats}
\usepackage{amsmath}
\usepackage{array}
\usepackage{amssymb}
\usepackage{makecell}
\usepackage{stfloats}
\usepackage{booktabs}
\usepackage{threeparttable}
\usepackage{multirow}
\usepackage{float}
\usepackage{dsfont}
\usepackage{enumerate}
\usepackage{cleveref}
\allowdisplaybreaks[4]

\normalsize

\ifCLASSINFOpdf
\else
\fi
\UseRawInputEncoding
\begin{document}
\title{Bandlimited Communication With One-Bit Quantization and Oversampling: Transceiver Design and Performance Evaluation}

\author{ Rui Deng, Jing~Zhou,~\IEEEmembership{Member,~IEEE}, and~Wenyi~Zhang,~\IEEEmembership{Senior~Member,~IEEE}
\thanks
{This work was supported by the National Key Research and Development Program of China (2018YFA0701603), the National Natural Science Foundation of China under Grant 61722114, Key Research Program of Frontier Sciences of CAS (QYZDY-SSW-JSC003), and China Postdoctoral Science Foundation (2019M662197) (Rui Deng and Jing Zhou are co-first authors).

The authors are with the CAS Key Laboratory of Wireless-Optical Communications, University of Science and Technology of China, Hefei, China (e-mail: ruideng@mail.ustc.edu.cn; jzee@ustc.edu.cn; wenyizha@ustc.edu.cn).

{Code available at: https://github.com/Ray-Deng/One-Bit-Quantization-and-Oversampling}
}
}

\maketitle

\begin{abstract}
We investigate design and performance of communications over the bandlimited Gaussian channel with one-bit output quantization.
A transceiver structure is proposed, which generates the channel input using a finite set of time-limited and approximately bandlimited waveforms, and performs oversampling on the channel output by an integrate-and-dump filter preceding the one-bit quantizer.
The waveform set is constructed based on a specific bandlimited random process with certain zero-crossing properties which can be utilized to convey information.
In the presence of the additive white Gaussian noise, a discrete memoryless channel model of our transceiver is derived.
Consequently, we determine a closed-form expression for the high signal-to-noise-ratio (SNR) asymptotic information rate of the transceiver, which can be achieved by independent and identically distributed input symbols.
By evaluating the fractional power containment bandwidth,
we further show that at high SNR, the achievable spectral efficiency grows roughly logarithmically with the oversampling factor,
coinciding with a notable result in the absence of noise \cite{Shamai1994}.
Moreover, the error performance of our transceiver is evaluated by low density parity check coded modulation.
Numerical results demonstrate that reliable communication at rates exceeding one bit per Nyquist interval can be achieved at moderate SNR.
\end{abstract}


\IEEEpeerreviewmaketitle

\section{Introduction}\label{sec_intro}
With the increasing demand for data rates, analog-to-digital converters (ADCs) at the receiver have been becoming a major bottleneck in many state-of-the-art communication systems \cite{SPM-bottleneck09,Fettweis19}.
High-speed and high-resolution (e.g., 8-12 bits) ADCs have negligible quantization effects, but are very costly.
Moreover, they consume a great deal of power which will further increase by two to four times if the resolution increases by one bit \cite{Walden1999,Murmann15}.
Under such circumstances, using low-resolution (1-3 bits) ADCs at the receiver (i.e., performing coarse quantization on the channel output) has attracted considerable attention in recent years, and the one-bit ADC, whose output is the sign of its input, is of particular interest due to its superiority in cost and power efficiency; see, e.g., \cite{Nossek,singh2009limits,zeitler2012low-precision,mo2015capacity,Zhang16,CMH16,
Studer16,Rini17ITW,Abbas17,Rini18ISIT}.
Reducing the resolution of the ADC, however, causes nonlinear distortion which degrades the performance.
At low to moderate signal-to-noise ratio (SNR), the performance degradation is usually acceptable \cite{Nossek,singh2009limits}.
For example, in the additive white Gaussian noise (AWGN) channel, a one-bit symmetric output quantizer causes approximately $2$ dB power loss at low SNR \cite{VO79}.\footnote{We note that at low SNR the loss can be further reduced by using carefully designed asymmetric one-bit quantizer and asymmetric signal constellations \cite{KochLapidoth}.
However, since the low SNR regime is not our focus,
we only consider the symmetric one-bit quantizer and refer to it as ``one-bit quantizer'' for brevity.}
At high SNR the information rate is limited by the resolution of the ADC so that the performance degradation becomes significant; see, e.g., \cite{mo2015capacity,Rini18ISIT}.

It should be noted that most existing studies on communications with low-resolution ADCs consider discrete-time memoryless channels, including both scalar and vector cases,
and thereby explicitly or implicitly assume intersymbol-interference (ISI)-free linear modulation (including multicarrier modulation which satisfies the generalized Nyquist criterion \cite{FU98}) at the transmitter and symbol-rate sampling at the receiver.
On this premise, the information rate (excluding degrees-of-freedom gains in multiantenna systems) is upper bounded by the resolution of the quantizer, i.e., $n$ bits per Nyquist interval ($2n$ bits/s/Hz or $n$ bits/dimension) if an $n$-bit quantizer is used, since the ISI-free symbol rate cannot exceed the Nyquist rate.
In fact, results in \cite{zeitler2012low-precision} imply that the limit holds no matter whether the ISI exists.
However, as shown in \cite{Gilbert1993}, in a bandlimited \emph{noiseless} channel with a one-bit output quantizer, the information rate under \emph{oversampling} exceeds one bit per Nyquist interval.
By constructing a specific bandlimited random process as the channel input, in \cite{Shamai1994} it was further proved that the information rate under the same setting grow logarithmically with the oversampling factor.
It is therefore desirable to investigate whether we can substantially increase the information rate in the presence of noise through oversampling.

In a strictly bandlimited AWGN channel with one-bit output quantization, it has been proved that oversampling outperforms Nyquist sampling in both the capacity per unit-cost \cite{KochLapidoth1} and the achievable rate at high SNR \cite{Zhang2012}.
Specifically, using Gaussian codebook, information rates slightly greater than one bit per Nyquist interval can be achieved by combining oversampling and nearest neighbor decoding \cite{Zhang2012}.
In a series of works \cite{Krone2012,Halsig2014,Singh2015,landauCL,landauJWCN,Bender,landauTWC}, the benefit of oversampling in continuous-time AWGN channels with one-bit output quantization was studied, where various types of channel inputs with different bandwidth properties have been proposed.
In \cite{Krone2012,Halsig2014,Singh2015,landauCL,landauJWCN} the channel inputs were constructed by linear modulation (not necessarily ISI-free) with various constellations (e.g., ASK, QAM, PSK), symbol rates, pulse functions (including strictly bandlimited ones and others), and particularly different kinds of sequences which boost the information rates achieved.
Some special nonlinear input designs were also proposed, including a waveform design with exponentially distributed zero-crossing distances (which carry the information) and bounded amplitude \cite{Bender}, and a continuous phase modulation (CPM) based scheme \cite{landauTWC}.
These works showed that when the SNR is sufficiently high, spectral efficiency comparable to that obtained in the noiseless case \cite{Shamai1994} can be achieved under the same effective oversampling factors (with respect to the fractional power containment bandwidth).
In \cite{Bender}, by deriving a capacity lower bound for the bandlimited AWGN channel with one-bit quantized continuous-time output, the limit of the performance when the oversampling factor grows without bound was studied.
Additionally, recent investigations on one-bit quantized massive multiple-input and multiple-output (MIMO) systems showed that considerable performance gains (in terms of spectral efficiency or SNR) can be obtained by oversampling \cite{GBLV17,UY18}.

Although several transceiver designs under one-bit output quantization have been proposed, and the benefit of oversampling has been demonstrated by analytical or numerical results, the way to practical systems still needs to be explored.
In fact, some proposed designs aimed at improving information rates are rather complex for implementation.
For example, in \cite{landauJWCN,landauCL,landauTWC},
the transmitters employ long sequences which are specifically optimized with respect to alphabets and symbol rates,
and due to the introduced channel memory the receivers must perform maximum-likelihood sequence detection to achieve the predicted information rates.
Until recently, even in simpler cases such as linear modulation with independently and uniformly distributed (IUD) input symbols, it is still unclear how to design coded systems to approach the promised information rates.
For example, in \cite{Halsig2014,Singh2015},
numerical results showed that acceptable error performances can be achieved using linear modulation with non-IUD inputs at the cost of increased detection complexity, but IUD inputs do not work well.
The complexity will be further increased if channel codes and practical pulse shaping filters are introduced.
The latest works \cite{Fettweis} and \cite{Alencar19} made some progress on coding design for sequence inputs and CPM inputs, respectively, with soft-input soft-output decoding under one-bit quantization and oversampling.

In this paper, we consider the continuous-time AWGN channel and propose a transceiver structure with one-bit quantization at the receiver, wherein:
i)
The transmitter maps independently and identically distributed (IID) input symbols to approximately bandlimited channel inputs through a set of finite-duration waveforms, which is constructed based on the bandlimited random process investigated in \cite{Shamai1994};
ii)
The receiver includes an integrate-and-dump filter and a one-bit quantizer, which generate multiple one-bit observations of the noisy signal waveform in each Nyquist interval.
Such a design enables simple signal detection.
Its achievable information rate is easy to evaluate.
Moreover, standard channel codes can be readily utilized, and we use low density parity check (LDPC) bit-interleaved coded modulation (BICM) as an example.
The optimization of the waveform set as well as the mapping rule from bits to waveforms is also studied.
Information theoretic results and error performance simulations demonstrate that oversampling indeed increases the information rate significantly in noisy channels,
and spectral efficiencies evidently larger than one bit per Nyquist interval can be achieved by standard coded modulation techniques.
At high SNR, the achievable rate grows approximately logarithmically with the oversampling factor, consistent with the results obtained in \cite{Shamai1994} in the absence of noise.
The integrate-and-dump filter \cite{DC} has been extensively used in practice; see, e.g., \cite{P}.
We note that it was also used in a previous work \cite{landauJWCN} (see literature review above), where the length of an interval of integration is a full Nyquist interval.
In our work, however, the integration is performed in sample-level, see Sec. II-B.

\subsection{A Further Discussion on Sampling and Nonlinearity}

In view of the fact that the topic of communications under one-bit quantization spans several different communities,
here we provide a further discussion on related literature to
place our work in a broader context, with emphasis on the \emph{theoretical} benefit of oversampling in the presence of nonlinearity.
In practice, of course, oversampling is useful for signal processing, no matter whether nonlinearity exists.

In linear Gaussian channels, regardless of the transmitter, the Nyquist sampling at the receiver provides sufficient statistics of the channel output in general, and hence no benefit can be obtained by oversampling.
Undersampling, on the other hand, {typically incurs a capacity penalty (see \cite{Eldar13}, which also suggested exceptions).
For a given linear modulation scheme in a linear Gaussian channel,
it is common to perform matched filtering followed by symbol-rate sampling, which also provides sufficient statistics.
Here, the symbol-rate sampling may be either undersampling or oversampling (discarding the ISI-free constraint),
but it leads to the same performance as the Nyquist sampling.

The situation becomes drastically different in channels with nonlinear distortion, where oversampling is generally beneficial.
From a frequency-domain perspective, nonlinearity typically results in bandwidth expansion (see, e.g., \cite{WTT77})
so that Nyquist sampling is in general suboptimal.\footnote{As an exception, a special case is that, if the nonlinearity can be completely compensated,
the input signal can be perfectly recovered from its Nyquist samples \cite{Eldar}. Of course, the effect of quantization cannot be compensated even in the noiseless case.}
This interpretation follows from the fact that,
although the nonlinearity (e.g., quantization) is performed after sampling, the whole process is equivalent to passing the received signal through a quantizer in continuous time followed by a sampler.
Alternatively, for the special case of one-bit output quantization treated in this paper, we have an intuitive time-domain interpretation of the benefit of oversampling:
the resolution of the zero-crossing positions of the received waveform signal can be improved by observing its sign process multiple times in each Nyquist interval.
As noted earlier, the benefit of oversampling in this case has been demonstrated in
\cite{Gilbert1993,Shamai1994,KochLapidoth1,Zhang2012,Krone2012,Halsig2014,Singh2015,landauCL,landauJWCN,Bender,landauTWC,GBLV17,UY18},
some of which had paid attention to variations of linear modulation schemes.
The above time-domain interpretation reveals a potential connection between achievable performance under one-bit output quantization and zero-crossing design under spectral constraints.\footnote{The problem of zero-crossings of stochastic processes is a classical topic; see [\ref{Wong}, Sec. II], \cite{Requicha80}, \cite{Shamai1994}, and references therein.}
Therefore,
the research on one-bit output quantization should focus not only on oversampling-based receivers, but also on a redesign of modulation mechanism (as noted in \cite{Fettweis} recently; see also related works \cite{MLR2,MLR1} pointed out by an anonymous reviewer), rather than maintaining the traditional design in linear channels.

Finally, we mention a few other related problems.
i) It is natural to ask whether oversampling also provides benefit in the case of multi-bit output quantization; we leave it to future research.
ii) Besides coarse output quantization, significant benefit of oversampling has also been shown in other channels, e.g., Wiener phase noise channels\cite{GK17}.
iii) Low resolution digital-to-analog converters (DACs) at the transmitter have also received some attention recently; see, e.g., \cite{Jacobsson17DAC,Dutta20}.
A corresponding problem, which does not exist in studies of low resolution ADC, is that the bandwidth expansion of the continuous-time transmitted signal should be regulated to meet the spectral mask constraint of the channel.

\emph{Organization}:
The remaining part of this paper is organized as follows.
In Sec. II, we briefly introduce the problem and propose our transceiver structure.
Detailed descriptions, performance analysis, and numerical results are provided in subsequent sections.
In Sec. III, we first introduce the bandlimited random process proposed in \cite{Shamai1994} as our starting point,
and then provide details of the construction of our channel input.
Focusing on the high-SNR regime, Sec. IV provides an asymptotic information rate result including an explicit design that achieves it.
In Sec. V we study the optimization of our system, and provide numerical results including information rate calculation and error performance.

\emph{Notation}:
For sequences or tuples we use boldface letters and for sets we use calligraphic letters (e.g., $\mathcal S$).
The Fourier transform of $f$ is denoted by $\hat{f}$.
The all-one tuple is denoted by $\mathbf 1$.
We use ${\mathsf Q}(x)$ to denote the Q function defined as ${\mathsf Q}(x):=\frac{1}{\sqrt{2\pi}}\int_x^\infty\exp\left(-\frac{t^2}{2}\right)\mathrm dt$.
For a bounded continuous function $f(t)$, the notations $f(t_0^+)>0$ and $f(t_0^+)<0$ mean that there exists a constant $\epsilon>0$ such that $f(t)>0, t\in(t_0,t_0+\epsilon)$ and $f(t)<0, t\in(t_0,t_0+\epsilon)$, respectively.

\section{System Model}\label{sec_fundam}
\subsection{Transceiver Structure}
Consider communications over the continuous-time Gaussian channel
\begin{equation}
Y(t)=X(t)+Z(t),\mspace{4mu} t\in\mathbb R,
\end{equation}
where $Z(t)$ is white Gaussian noise of double sided power spectral density (PSD) $ N_0/2$.
The channel input $X(t)$ is limited by a fractional power containment bandwidth $W_\eta$, which satisfies
\begin{equation}
\label{Weta}
\int_{-W_\eta}^{W_\eta} \mathsf{S}(f) \mathrm df=\eta \mathsf P,
\end{equation}
where $\mathsf{S}(f)$ is the PSD of $X(t)$,\footnote{We use the definition in [\ref{Lapidoth2017}, Definition 15.3.1], because it holds for cyclostationary random process. It can be calculated by taking the Fourier transform of the average autocovariance function of $X(t)$. In \cite{Lapidoth2017} it is called the operational PSD.} $\eta$ is a given constraint smaller than (but typically close to) one, and $\mathsf P$ is the power of $X(t)$:
\begin{equation}
\mathsf P:=\lim\limits_{T\to\infty}\frac{1}{2T}\mathrm E\left[\int_{-T}^{T}X^2(t)\mathrm dt\right]= \int_{-\infty}^{\infty} \mathsf{S}(f) \mathrm df.
\end{equation}
Our basic assumption is that a one-bit quantizer, instead of a high-resolution one, is applied at the receiver.
As shown in Fig. \ref{fig:a}, we propose a transceiver which performs oversampling before the one-bit quantizer to help recover the transmitted information.
In the following, we first briefly introduce our design, and details will be given in subsequent sections.

\begin{figure} \centering
\subfigure[Proposed transceiver.] { \label{fig:a}
\includegraphics[width=1\columnwidth]{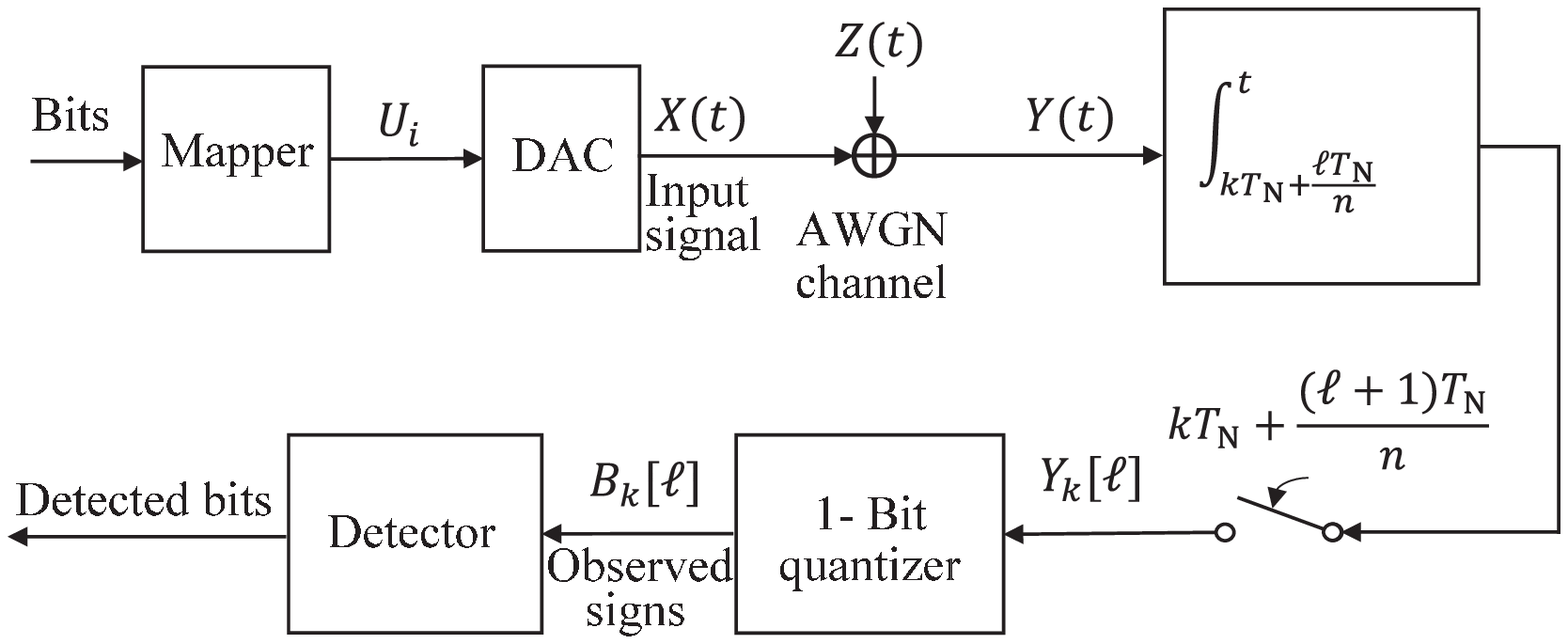}
}
\subfigure[{Equivalent transceiver of \cite{Shamai1994}.}] { \label{fig:b}
\includegraphics[width=1\columnwidth]{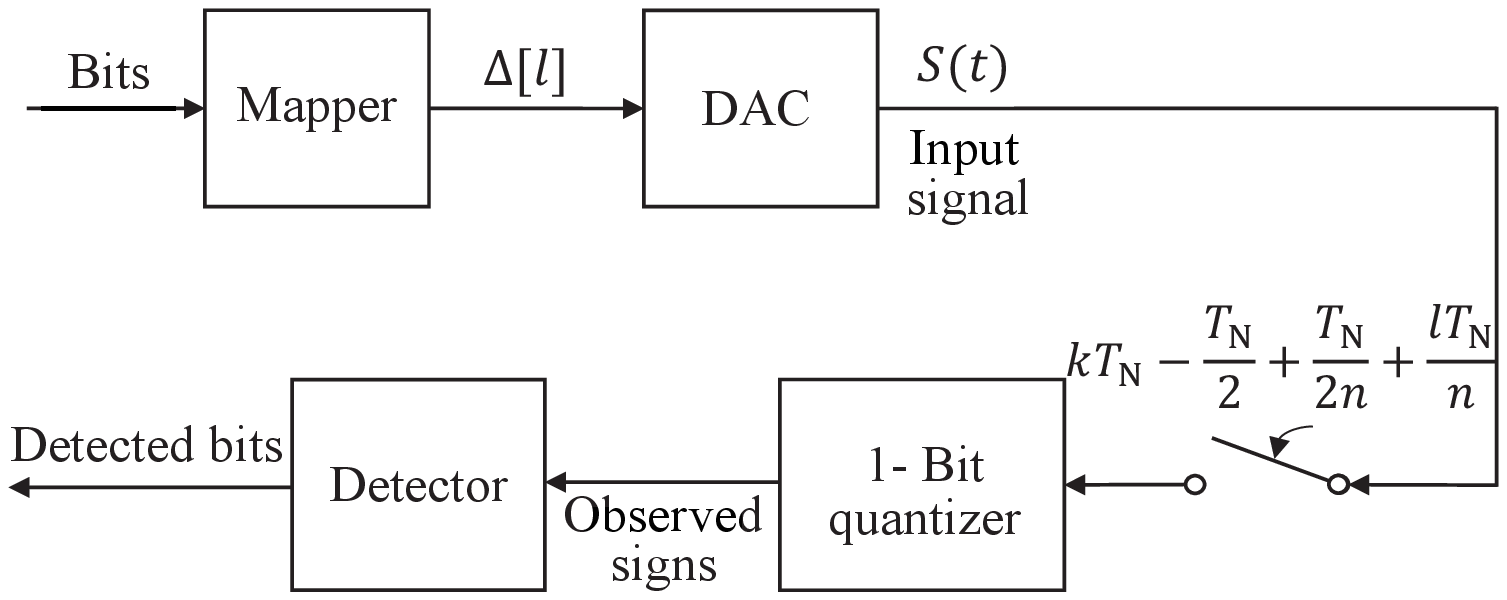}
}
\caption{System model.}
\label{fig_system}
\end{figure}

\subsubsection{Transmitter}
We construct $X(t)$ based on a bandlimited random process $S(t)$ with certain zero-crossing properties (see Sec. \ref{subsec_basis_construction}).
The process was proposed in \cite{Shamai1994} for communications with one-bit quantization and oversampling at the receiver in the absence of noise.
Due to its infinite duration, $S(t)$ cannot be used as the channel input directly.
Instead, we use truncations of realizations of $S(t)$ as building blocks of $X(t)$, where the realizations are chosen carefully to ensure distinguish ability after one-bit quantization at the receiver if the noise is absent.
After some amplitude- and time-scaling (see Sec. \ref{sec_construction}), we obtain a set ${\mathcal G}=\{g_u(t),u=1,...,m\}$ including $m$ finite-duration and finite energy waveforms satisfying $g_u(t)=0$ for $t\in(-\infty,0)$ and $t\in(\kappa T_{\mathrm N},\infty),\kappa\in \mathbb{N}$,
where $T_{\mathrm N}$ is the length of a Nyquist interval which contains exactly one zero-crossing.
Based on $\mathcal G$, the channel input $X(t)$ is given by
\begin{align}
\label{Xt}
X(t) = \sum_{i}G_{U_i}(t-i\kappa T_{\mathrm N}),
\end{align}
where $G_{U_i}(t)$ is a finite-duration random waveform given by
\begin{align}
\label{G}
G_{U_i}(t) = \sum_{u=1}^{m}\mathds{1}_{U_i}(u)g_u(t),
\end{align}
where $\mathds{1}_{U_i}(u)$ is an indicator defined as
\begin{equation}
\label{U}
\mathds{1}_{U_i}(u):=\left\{
\begin{aligned}
1, & & {u=U_i}&\\
0, & & {u\neq U_i}&,
\end{aligned} \right.
\end{equation}
and $\{U_i\}$ is a sequence of IID random variables distributed over $\mathcal U=\{1,...,m\}$.
Therefore, the channel input $X(t)$ can be generated by
i) a digital mapper, which maps the information bits to $\{U_i\}$,
and ii) a deterministic and memoryless DAC, which maps $\{U_i\}$ to $X(t)$ according to (\ref{Xt}) and (\ref{G}).
The SNR is defined as $\mathsf{SNR}:=\frac{\mathsf P}{N_0W_{\mathrm N}}$. We call $W_{\mathrm N}=\frac{1}{2 T_{\mathrm N}}$ the \emph{nominal bandwidth} of $X(t)$, which corresponds to the bandwidth of $S(t)$ through the time-scaling.
Due to truncation, the fractional power containment bandwidth of $X(t)$, namely $W_\eta$, is slightly larger than $W_{\mathrm N}$.
More details are given in subsequent sections.

\subsubsection{Receiver}
The received signal $Y(t)$ is uniformly sampled $n$ times in each Nyquist interval (i.e., the oversampling factor with respect to $W_{\mathrm N}$ is $n$),
yielding a sequence of $n$-tuples $\{\mathbf{Y}_{k}\}$, $k\in \mathbb{Z}$, where the $n$-tuple $\mathbf{Y}_{k}=\left[Y_k[1],...,Y_k[n]\right]$ corresponds to the $k$-th Nyquist interval.
In contrast to the noiseless case in Fig. \ref{fig:b}, a key additional step is an integrator before the sampler.
This receiver structure, although in general suboptimal, enables simple signal detection and performance evaluation in the presence of AWGN and can lead to asymptotically optimal performance at high SNR (see Sec. IV).
The output of the one-bit quantizer in the $k$-th Nyquist interval, denoted as $\mathbf{B}_{k}=\left[B_k[1],...,B_k[n]\right]$, is an $n$-tuple with binary elements satisfying ${B}_k[\ell]=\mathrm{sgn}\left(Y_k[\ell]\right)$, $1\leq\ell\leq n$, where\footnote{Following \cite{Shamai1994}, in our definition we let $\mathrm{sgn}(\cdot)$ be binary for simplicity, although defining $\mathrm{sgn}(0)=0$ is more common. }
\begin{equation}
\mathrm{sgn}(y):=\left\{
\begin{aligned}
1, & & y\ge 0&\\
-1, & & y<0&.
\end{aligned} \right.
\end{equation}
The task of the detector in Fig. \ref{fig:a} is to recover the information from $\{\mathbf{B}_{k}\}$ with a sufficiently low error probability.
The combination of the integrator and the sampler is also known as an \emph{integrate-and-dump} filter \cite{DC}, which resets the integrator to zero periodically.
For signal detection, the additive noise may add or erase zero-crossings which complicate the detection, but the integrate-and-dump filter is not very sensitive to such noise-added/erased zero-crossings.
Furthermore, the integrate-and-dump filter leads to independent observations on the received noisy signal in intervals of integration, thereby yielding a discrete-time memoryless channel model for receiver design and performance analysis.
\subsection{A Discrete Memoryless Channel Model}
At our receiver, the output of the integrate-and-dump filter is
\begin{align}
\label{YXZ}
Y_k[\ell]&=\int_{(k-1+\frac{\ell-1}{n})T_\mathrm{N}}^{(k-1+\frac{\ell}{n})T_\mathrm{N}}Y(t)\mathrm dt\notag\\
&=\int_{(k-1+\frac{\ell-1}{n})T_\mathrm{N}}^{(k-1+\frac{\ell}{n})T_\mathrm{N}}X(t)\mathrm dt+\int_{(k-1+\frac{\ell-1}{n})T_\mathrm{N}}^{(k-1+\frac{\ell}{n})T_\mathrm{N}}Z(t)\mathrm dt\notag\\
&=X_k[\ell]+Z_k[\ell], \mspace{4mu}\ell=1,...,n,\mspace{4mu}k=1,2,...
\end{align}
where $Z_k[\ell]$ is IID satisfying $Z_k[\ell]\sim \mathcal N(0,\frac{N_0T_\textrm{N}}{2n})$ \cite{Lapidoth2017}.
It should be noted that if we perform oversampling directly on $Y(t)$, we cannot obtain such an IID noise term.
The output of the one-bit quantizer is
\begin{align}
\label{onebit}
{B}_k[\ell]=\mathrm{sgn}\left(Y_k[\ell]\right)=\mathrm{sgn}\left({X}_k[\ell]+Z_k[\ell]\right),\mspace{4mu}\ell=1,...,n.
\end{align}
From (\ref{Xt})-(\ref{onebit}), the resulting discrete-time channel is \emph{memoryless} since the current output depends only on one symbol in the input sequence $\{U_i\}$ (note that there is no overlap between successive waveforms $G_{U_i}(t-i\kappa T_{\mathrm N})$ in (\ref{Xt})).
Hence we can omit the index $i$ and consider the mapping from an input symbol $U$ to the corresponding noiseless samples (a length-${\kappa}$ sequence of $n$-tuples) at the receiver.
When $U=u$, the channel input in $(0,\kappa T_\mathrm N]$ is $X(t)=g_u(t)$, where we assume that the transmission begins at $t=0$ without loss of generality. The output of the integrate-and-dump filter is thus
\begin{align}
\label{GGZ}
Y_{k}[\ell]&=\int_{(k-1+\frac{\ell-1}{n})T_\mathrm{N}}^{(k-1+\frac{\ell}{n})T_\mathrm{N}}g_{u}(t)dt+\int_{(k-1+\frac{\ell-1}{n})T_\mathrm{N}}^{(k-1+\frac{\ell}{n})T_\mathrm{N}}Z(t)dt\notag\\
&=g_{u,k}[\ell]+Z_k[\ell], \mspace{4mu}\ell=1,...,n,\mspace{4mu}k=1,...,\kappa,
\end{align}
and we can rewrite it as
\begin{align}
\label{fu}
{\mathbf Y}^\kappa=\mathbf{g}_{u}^\kappa+\mathbf{Z}^\kappa,
\end{align}
where $\mathbf{g}_{u}^\kappa=[\mathbf{g}_{u,1},...,\mathbf{g}_{u,\kappa}]$, $\mathbf{g}_{u,k}=[g_{u,k}[1],...,g_{u,k}[n]]$, and similar notations also apply for the noise term.
Consequently, a discrete-time memoryless channel model of our transceiver is obtained as
\begin{align}
\label{DMC}
\mathbf{B}^{\kappa}=\mathsf{sgn}\left(\mathbf{G}^\kappa+\mathbf{Z}^{\kappa}\right),
\end{align}
where $\mathsf{sgn}(\mathbf A):=[\mathrm{sgn}(A_1),...,\mathrm{sgn}(A_k)]$, and
\begin{align}
\mathbf{G}^\kappa=\sum_{u=1}^{m}\mathds{1}_{U}(u)\mathbf{g}_{u}^\kappa.
\end{align}

Clearly, (\ref{DMC}) is a discrete memoryless channel (DMC) with an input alphabet $\mathcal U=\{1,..,m\}$, an output alphabet $\mathcal B=\{1,-1\}^{\kappa n}$, and transition probabilities $p_{\mathbf{B}^{\kappa}|U}(\mathbf{b}^{\kappa}|u)$\footnote{For brevity we write $p (\mathbf{b}^{\kappa}|u)$ hereinafter. Similarly, for the input distribution $p_U(u)$ we write $p(u)$.}
which can be evaluated as
\begin{align}
p \left(\mathbf{b}^{\kappa}|u\right)&=\prod\limits_{k=1}^\kappa\prod\limits_{\ell=1}^np\left(b_k[\ell]|g_{u,k}[\ell]\right)\notag\\
&=\prod\limits_{k=1}^\kappa\prod\limits_{\ell=1}^n\Pr\left(\mathrm{sgn}(g_{U,k}[\ell]+Z_k[\ell])=b_k[\ell]|U=u\right).
\end{align}
From
\begin{align}
&\Pr\left(\mathrm{sgn}\left(g_{U,k}[\ell]+Z_k[\ell]\right)=1|U=u\right)\notag\\
=&\Pr\left(g_{u,k}[\ell]+Z_k[\ell]>0\right)\notag\\
=&\Pr\left(Z_k[\ell]>-g_{u,k}[\ell]\right)\notag\\
=&\mathsf Q\left(-\sqrt{\frac{2n}{N_0T_\mathrm{N}}}g_{u,k}[\ell]\right),
\end{align}
\begin{align}
&\Pr\left(\mathrm{sgn}\left(g_{U,k}[\ell]+Z_k[\ell]\right)=-1|U=u\right)\notag\\
=&1- \mathsf Q\left(-\sqrt{\frac{2n}{N_0T_\mathrm{N}}}g_{u,k}[\ell]\right)\notag\\
=&\mathsf Q\left(\sqrt{\frac{2n}{N_0T_\mathrm{N}}}g_{u,k}[\ell]\right),
\end{align}
we have
\begin{align}
\label{pyx}
p \left(\mathbf{b}^{\kappa}|u\right)=\prod\limits_{k=1}^\kappa\prod\limits_{\ell=1}^n \mathsf Q\left(-\sqrt{\frac{2n}{N_0T_\mathrm{N}}}b_k[\ell]g_{u,k}[\ell]\right).
\end{align}

Since we use $\kappa$ Nyquist intervals to transmit an input symbol $U$,
the achievable information rate of our transceiver can be evaluated by the mutual information between the channel input and the channel output of the DMC (\ref{DMC}), in a closed form as
\begin{align}
\label{IR}
R&=\frac{1}{\kappa T_\mathrm{N}}I(U;\mathbf{B}^{\kappa})\notag\\
&=\frac{1}{\kappa T_\mathrm{N}}\sum\limits_{\mathbf{b}^\kappa\in\{0,1\}^{\kappa n}}\sum\limits_{u=1}^{m}p(u)p(\mathbf{b}^{\kappa}|u)\notag\\
&\qquad\qquad\qquad\qquad\qquad\log\frac{p(\mathbf{b}^{\kappa}|u)}{\sum_{u'=1}^{m}p(u')p(\mathbf{b}^{\kappa}|u')},
\end{align}
and the corresponding spectral efficiency is $\mathsf{SE}=\frac{R}{W_\eta}$.
According to the fact
\begin{equation}
I(U;\mathbf{B}^{\kappa})=H(\mathbf{B}^{\kappa})-H(\mathbf{B}^{\kappa}|U) \leq H(\mathbf{B}^{\kappa}),
\end{equation}
the information rate can be upper bounded by the entropy of the output of the one-bit quantizer.
Under Nyquist sampling ($n=1$), the information rate cannot exceed one bit per Nyquist interval.
We will show that the information rate can grow approximately logarithmically with the oversampling factor $n$, at high SNR,
coinciding with the noiseless case result of \cite{Shamai1994}.

\section{Construction of Channel Input}\label{sec_construction}
In this section, based on a bandlimited process proposed in \cite{Shamai1994},
we give details of the construction of the waveform set $\mathcal G$, which provides all $m$ waveforms $g_u(t)$, $u=1,...,m$ in (\ref{G}) so that the channel input $X(t)$ can be generated according to the transmitted symbol $U_i$.
\subsection{A Bandlimited Process With Certain Zero-Crossing Properties}\label{subsec_basis_construction}
In \cite{Shamai1994}, a bounded continuous random process
\begin{equation} \label{construction}
S(t)=(t-\tau_{0})\lim_{K\to\infty}\prod\limits_{k=1}^{K}\left(1-\frac{t}{\tau_{k}}\right)\left(1-\frac{t}{\tau_{-k}}\right)
\end{equation}
is considered, where $\{\tau_k, k\in \mathbb Z\}$ is a sequence of random variables satisfying $k-\frac{1}{2}\leq \tau_{k} < k+\frac{1}{2}$.
Actually, the sequence $\{\tau_k\}$ corresponds to all zero-crossings of $S(t)$.
In \cite{Shamai1994}, it was shown that $S(t)$ is of bandwidth $1/2$.\footnote{Here the bandwidth limitation is in the sense of Zakai \cite{Zakai1965}.}
Thus, the length of a Nyquist interval is one, and there is exactly one zero-crossing in each Nyquist interval.
We can scale $S(t)$ in time by a factor of $T_{\mathrm{N}}$, and the bandwidth of $S(t)$ becomes $W_\mathrm{N}=\frac{1}{2T_{\mathrm{N}}}$ accordingly.
A notable fact is that the sign of $S(t)$ at instants $t=k-\frac{1}{2}$ is known \textit{a prior} and alternates, that is,
\begin{equation}
\label{sign}
\mathrm {sgn}\left(S\left(k-\frac{1}{2}\right)\right)=(-1)^{k-1}.
\end{equation}
By varying the positions of zero-crossings in Nyquist intervals of $S(t)$, information can be conveyed.
An example of $S(t)$ was provided in \cite{Shamai1994} by letting $\{\tau_k-k\}$ be an IID sequence satisfying
\begin{equation}
\label{tau_s}
\Pr\left(\tau_k=k-\frac{1}{2}+\Delta[l]\right)=\frac{1}{n+1},\mspace{4mu} l=0,1,...,n,
\end{equation}
where
\begin{equation}
\label{pattern0}
\Delta[l]=
\left\{
\begin{aligned}
&\frac{1}{4n}, &&l=0\\
&\frac{l}{n}, &&l=1,...,n-1\\
&1-\frac{1}{4n}, &&l=n.
\end{aligned} \right.
\end{equation}
The set $\{\Delta[l]\}$, which indicates all possible zero-crossing positions in a Nyquist interval,
is referred to as a \emph{zero-crossing pattern} hereinafter.
In the noiseless case, by observing the signs of $S(t)$ at $t=k-\frac{1}{2}+\frac{l+0.5}{n},\mspace{4mu} l=0,1,...,n-1$ via oversampling,
the zero-crossings can be located perfectly, since there is exactly one sample point between each two possible zero-crossing positions.
See Fig. \ref{visio_sub_fig1}.
Therefore, the information conveyed by $\{\tau_k\}$ can be recovered, achieving an information rate $\log_2(n+1)$ bits per Nyquist interval \cite{Shamai1994},\footnote{
It was also proved in \cite{Shamai1994} that if the sign process $\mathrm{sgn}\left(S(t)\right)$ is stationary (the zero-crossing pattern (\ref{pattern0}) satisfies this condition), the achievable information rate cannot exceed $\log_2 n+(n-1)\log_2\left(\frac{n}{n-1}\right)$ bits per Nyquist interval when $n\ge 2$.
} which is equal to the entropy of $\tau_k$ in (\ref{tau_s}).
As a result, the information rate can be substantially improved through oversampling.
In the following, we turn to the noisy case.
\subsection{Two New Zero-Crossing Patterns}\label{subsec_Equally_Construction}
\begin{figure}
\includegraphics[width=2.8in,height=2.0in]{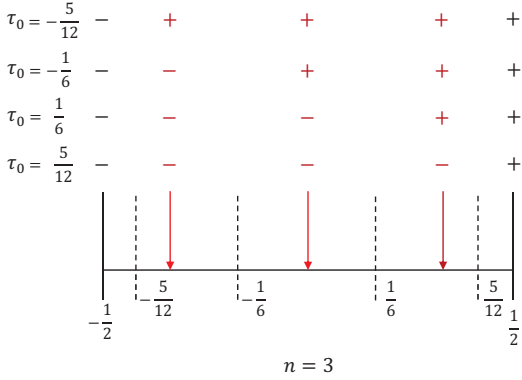}
\caption{Sampling points and zero-crossing positions of $S(t)$. }
\label{visio_sub_fig1}
\end{figure}
\begin{figure}
\includegraphics[width=2.8in,height=2.0in]{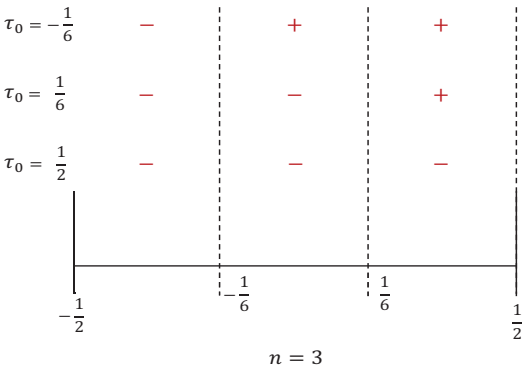}
\caption{Zero-crossings and integration intervals (separated by vertical lines) under uniform zero-crossing pattern.}
\label{visio_sub_fig2}
\end{figure}

In consideration of the integrate-and-dump filter at the receiver,
we introduce two new zero-crossing patterns for $S(t)$ to construct $X(t)$.
The new patterns satisfy $k-\frac{1}{2}<\tau_{k} \leq k+\frac{1}{2}$.
This is slightly different to the assumption in \cite{Shamai1994}, but it does not change the properties of $S(t)$.
\subsubsection{A Uniform Zero-Crossing Pattern}
Since the receiver only observes the sign of $Y_k[\ell]$, for reducing the error probability caused by noise, intuitively, we expect the sign of $S(t)$ to remain unchanged in each interval of integration.
To this end, a natural solution is to let the zero-crossings occur only at a sampling point (i.e., at the end of an interval of integration). This leads to a uniform zero-crossing pattern as
\begin{equation}
\label{tl}
\Delta[l]=\frac{l}{n},\mspace{4mu} l=1,...,n.
\end{equation}
See Fig. \ref{visio_sub_fig2}.
However, since there are $n$ possible zero-crossing positions in a Nyquist interval, the achievable information rate is upper bounded by $\log_2 n$ bits per Nyquist interval,
which is lower than that achieved by the example given by (\ref{tau_s}) and (\ref{pattern0}).

\subsubsection{A Nonuniform Zero-Crossing Pattern}
By allowing an extra zero-crossing position inside the first interval of integration of each Nyquist interval, we can design a zero-crossing pattern with $n+1$ possible zero-crossing positions, and still recover the information with a low error probability reliably at high SNR.
Let us employ a zero-crossing pattern as
\begin{equation} \label{tau_Unequally}
\Delta[l]=\left\{
\begin{aligned}
\frac{\lambda}{n}, &　&l=&0\\
\frac{l}{n}, &　&l=&1,...,n,
\end{aligned} \right.
\end{equation}
where $0<\lambda<1$.
Without loss of generality, consider the first Nyquist interval, in which we have ${\mathrm{sgn}} \left(\int_{\frac{1}{2}}^{\tau_1}S(t)\mathrm dt \right)=1$.
If $\tau_1=\frac{1}{2}+\Delta[0]$, the first sample at this Nyquist interval is
\begin{equation}
S_1[1]=\underbrace{\int_{\frac{1}{2}}^{\frac{1}{2}+\frac{\lambda}{n}}S(t)\mathrm dt}\limits_{>0}+\underbrace{\int_{\frac{1}{2}+\frac{\lambda}{n}}^{\frac{1}{2}+\frac{1}{n}}S(t)\mathrm dt.}\limits_{<0}
\end{equation}
By letting $\lambda$ be small enough, we can make the sum $S_1[1]$ negative.\footnote{Following the definition (\ref{construction}), it can be shown that $S_1[1]$ is a strictly increasing function of $\lambda$, and $\lim\limits_{\lambda\to 0}S_1[1]<0$. Hence such a positive $\lambda$ always exists.
}
When $\tau_1=\frac{1}{2}+\Delta[l]$, $l\ge 1$, $S_1[1]$ is positive.
Therefore, the sign of the first sample $\mathrm{sgn}(S_1[1]+Z_1[1])$ can be utilized to detect whether a zero-crossing occurs at the position corresponding to $\Delta[0]$.
In contrast, when using the uniform zero-crossing pattern as (\ref{tl}), the sign of the first sample is deterministic: $\mathrm{sgn}(S_k[1])=(-1)^{k-1}$, $k\in\mathbb{Z}$.

To further explain the two zero-crossing patterns, we take the case $n=4$ for example.
For even $k$, using the uniform zero-crossing pattern, the sign tuple for $\mathbf{S}_k$ has four possible realizations:
\begin{equation}
\label{ssku}
\mathsf{sgn}(\mathbf{s}_k)\in\left\{[- + + +], [- - + +], [- - - +], [- - - -]\right\},
\end{equation}
where for brevity we use $+$ and $-$ to represent $+1$ and $-1$, respectively.
Using the nonuniform zero-crossing pattern, the first sample $S_k[1]$ is no longer deterministic, and we have
\begin{align}
\label{sskn}
\mathsf{sgn}(\mathbf{s}_k)\in&\left\{[+ + + +], [- + + +], \right. \notag\\
&\qquad\qquad\left. [- - + +], [- - - +], [- - - -]\right\}.
\end{align}
When $k$ is odd, all the signs in the tuples in (\ref{ssku}) and (\ref{sskn}) should be flipped.
\subsection{Construction of Waveform Set $\mathcal G$ via Truncation}\label{sec_des_set}
As the process $S(t)$ has infinite duration and doubly infinite (noncausal) memory,
we employ $X(t)$ given in Sec. II-A as our channel input, where the waveform set $\mathcal G$ contains truncations of realizations of $S(t)$.
Clearly, the zero-crossing properties of $S(t)$ can still be utilized if we let the truncation interval contain an integer number of Nyquist intervals, i.e., let the truncation interval be $[k_0-\frac{1}{2},k_0+\kappa-\frac{1}{2}]$.
Then a waveform in $\mathcal G$ is obtained by
\begin{equation}
\label{gs}
g_u(t)=\varphi_u s\left(\frac{t+k_0-\frac{1}{2}}{T_{\mathrm N}}\right)\cdot\mathds{1}_{\left(0,\kappa T_\mathrm N\right]}(t),
\end{equation}
where $\varphi_u$ is an amplitude-scaling factor, $s(t)$ a realization of $S(t)$ determined by some specific zero-crossings, and $\mathds{1}_{(t_1,t_2]}(t)$ an indicator defined as
\begin{equation}
\label{truncation}
\mathds{1}_{(t_1,t_2]}(t):=\left\{
\begin{aligned}
1, & &t&\in(t_1,t_2]\\
0, & &t&\notin(t_1,t_2].
\end{aligned} \right.
\end{equation}
In numerical calculations, the interval $[k_0-\frac{1}{2},k_0+\kappa-\frac{1}{2}]$ of $s(t)$ in (\ref{gs}) can be approximated from the right hand side (RHS) of (\ref{construction}) with enough accuracy by removing the limit operation and letting $K$ be sufficiently large.
By varying the positions of the $\kappa$ zero-crossings in $(0, \kappa T_\mathrm N]$, we obtain multiple realizations of $S(t)$, each corresponding to a truncated waveform.
Specifically, under the uniform zero-crossing pattern we obtain $n^\kappa$ waveforms in total, and
under the nonuniform zero-crossing pattern we obtain $(n+1)^\kappa$ waveforms in total.
In Fig. \ref{figUZC} and Fig. \ref{fignUZC},
we show examples of waveforms obtained by the above truncation procedure, where we set $k_0=0$, $n=4$, and $\kappa=3$.

\begin{figure*}
\centering
\subfigure[{{An example of waveform corresponding to $\tau_0=-\frac{1}{2}+\Delta[1], \tau_1=1-\frac{1}{2}+\Delta[3], \tau_2=2-\frac{1}{2}+\Delta[2]$.}}]
{
\begin{minipage}[t]{0.48\linewidth} \centering \label{UZC_0}
\includegraphics[width=1\columnwidth]{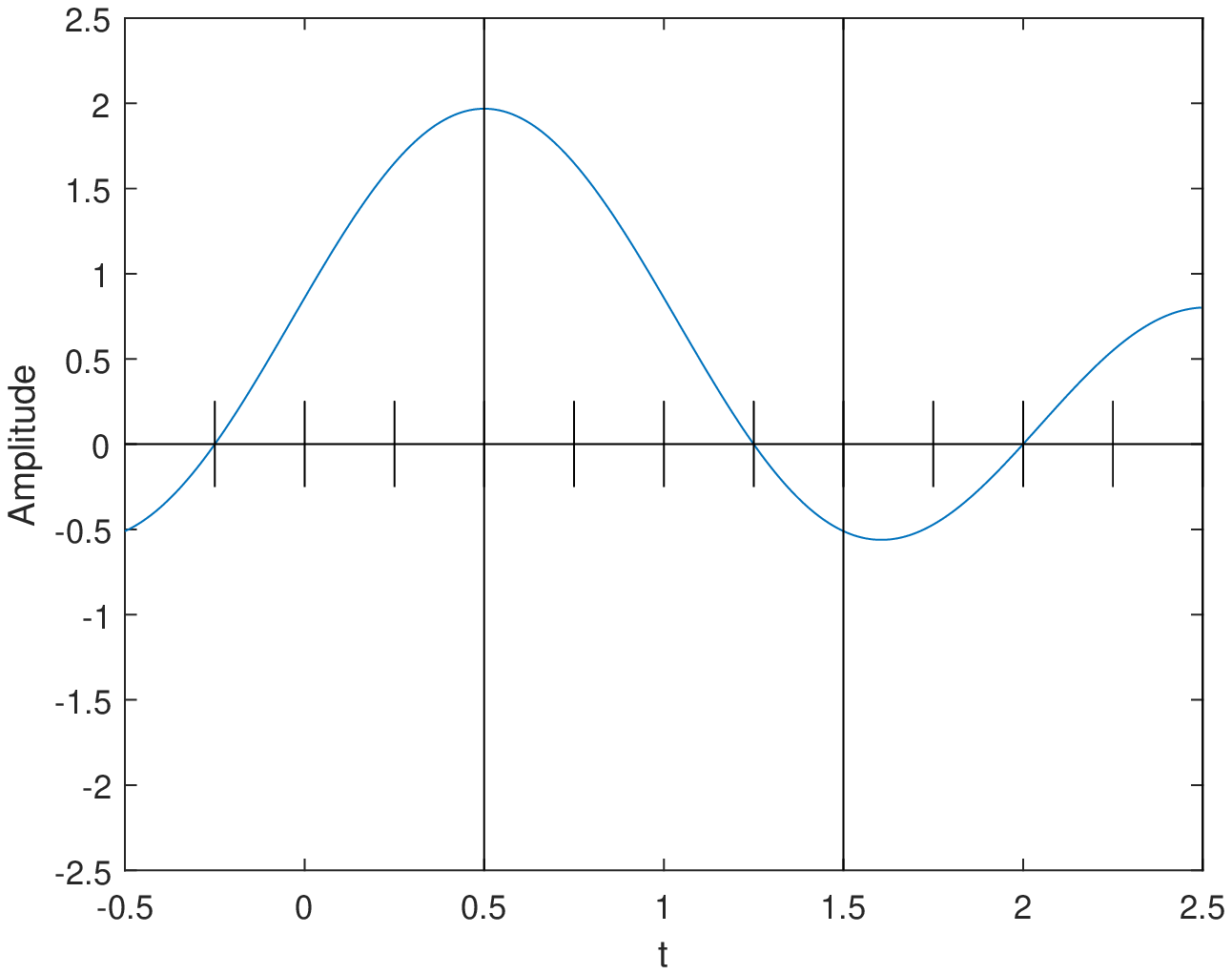}
\end{minipage}
}
\subfigure[All possible waveforms: $n^\kappa=4^3=64$ in total.]
{
\begin{minipage}[t]{0.48\linewidth} \centering \label{UZC}
\includegraphics[width=1\columnwidth]{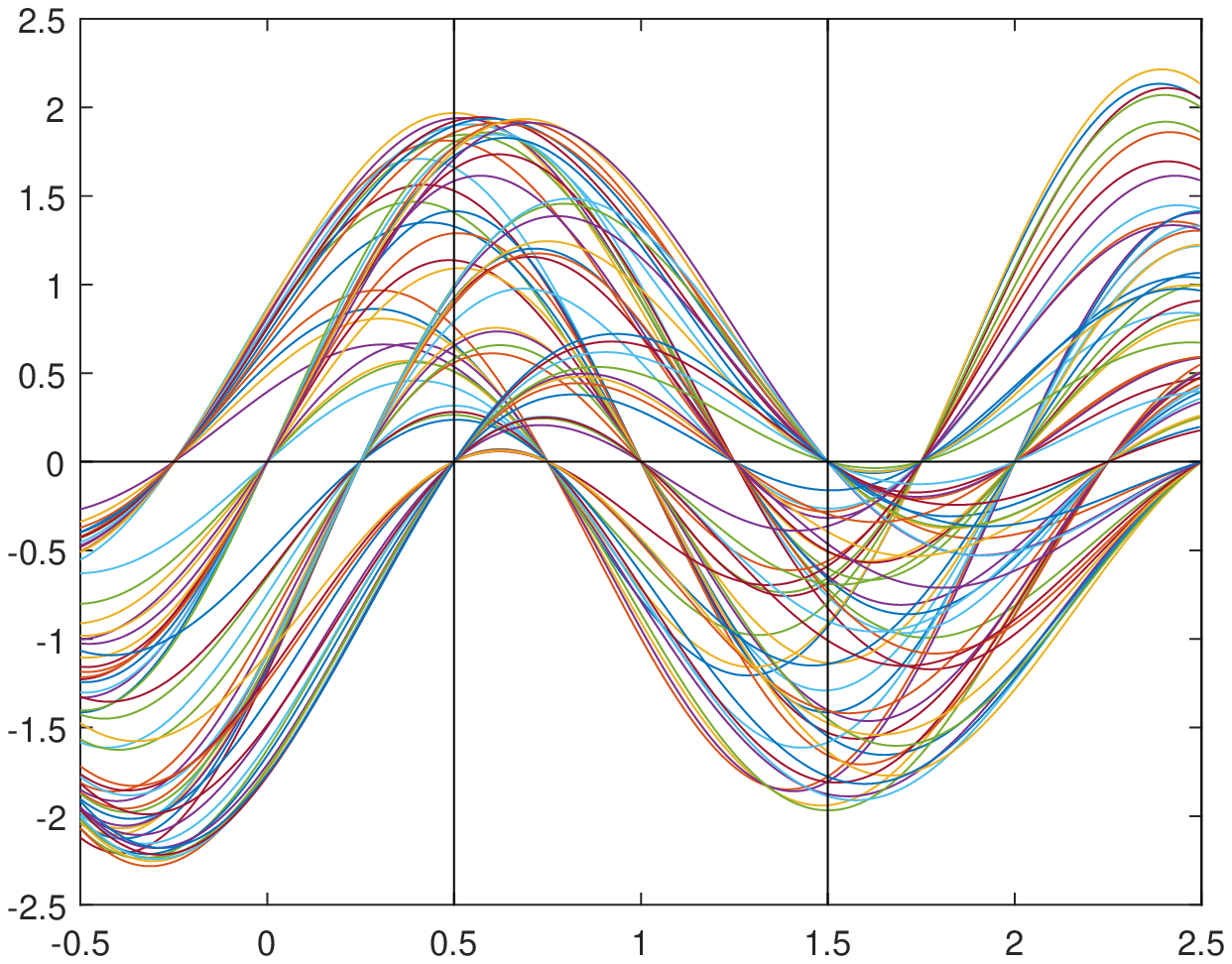}
\end{minipage}
}
\caption{Waveforms obtained by truncating realizations of $S(t)$ with the uniform zero-crossing pattern.}
\label{figUZC}

\centering
\subfigure[{{An example of waveform corresponding to $\tau_0=-\frac{1}{2}+\Delta[0], \tau_1=1-\frac{1}{2}+\Delta[1], \tau_2=2-\frac{1}{2}+\Delta[2]$}}.]
{
\begin{minipage}[t]{0.48\linewidth} \centering \label{nUZC_0}
\includegraphics[width=1\columnwidth]{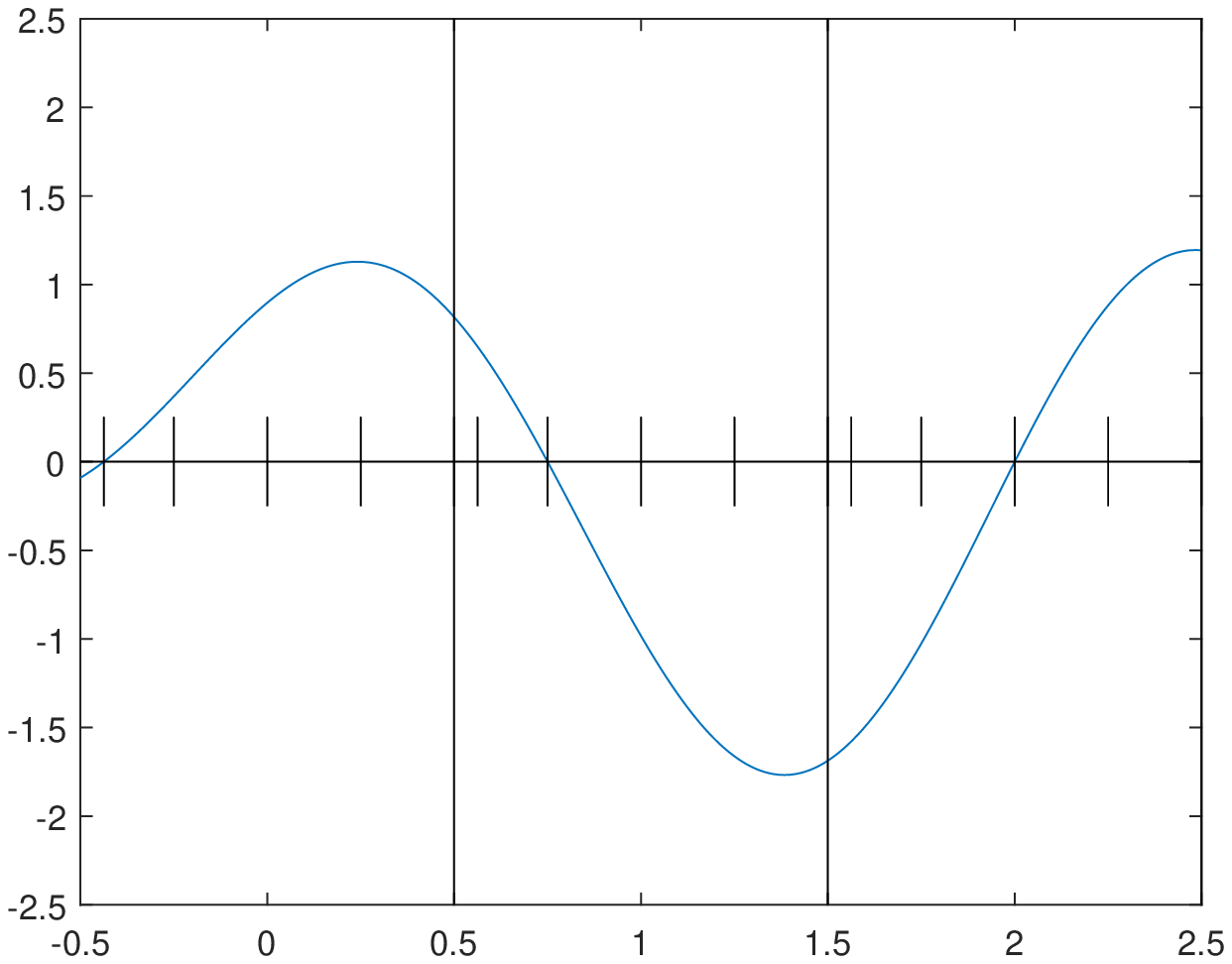}
\end{minipage}
}
\subfigure[All possible waveforms: $(n+1)^\kappa=5^3=125$ in total.]
{
\begin{minipage}[t]{0.48\linewidth} \centering \label{nUZC}
\includegraphics[width=1\columnwidth]{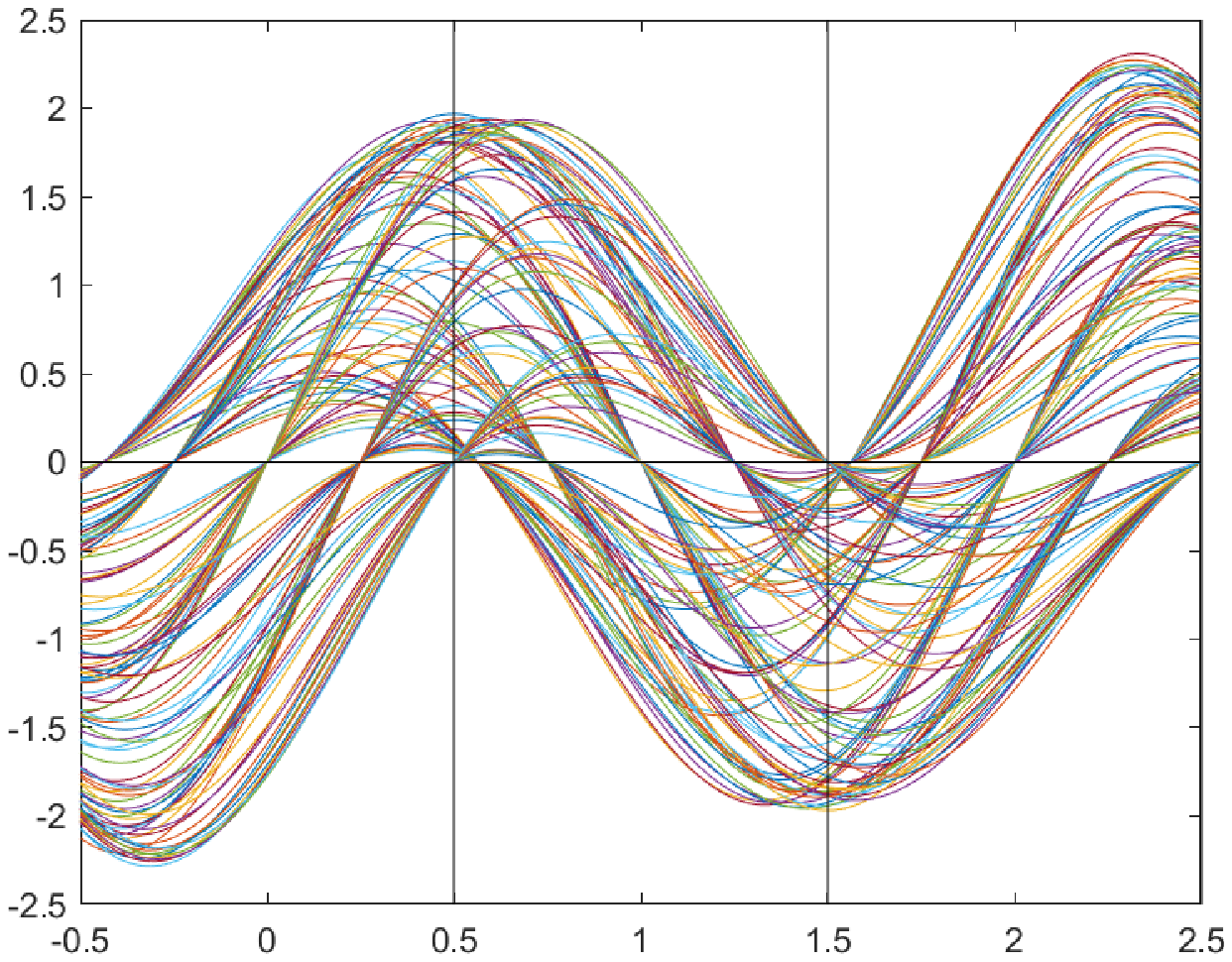}
\end{minipage}
}
\caption{Waveforms obtained by truncating realizations of $S(t)$ with the nonuniform zero-crossing pattern ($\lambda=1/4$).}
\label{fignUZC}
\end{figure*}

Next, we discuss various aspects in the construction of $\mathcal G$ as follows.
\subsubsection{A Uniqueness Condition}
To convey more information for a given $\kappa$, we prefer a larger $|\mathcal G|$.
Although more waveforms can always be constructed by increasing the number of possible zero-crossing positions of $S(t)$,
at least we must avoid ambiguity at the receiver when the noise is absent.
Therefore we impose a \emph{uniqueness condition} in the construction of $\mathcal G$:
\emph{when the noise is absent, for a given channel output} $\mathbf{b}^{\kappa} \in \{1,-1\}^{\kappa n}$, \emph{the corresponding input waveform in} $\mathcal G$, \emph{if exists, is unique}.
In the DMC (\ref{DMC}), when $U=u$, the channel output in the absence of noise is $\mathbf b^\kappa=\mathsf{sgn}(\mathbf{g}_{u}^\kappa)$, which is a length $\kappa$ sequence of $n$-tuples.
We call $\mathsf{sgn}(\mathbf{g}_{u}^\kappa)$ the \emph{sign sequence} of $g_u(t)$, and we use $\mathcal S_\mathcal G$ to denote the set of the sign sequences of waveforms in $\mathcal G$, i.e., ${\mathcal S}_{\mathcal G}=\{\mathsf{sgn}(\mathbf{g}_{u}^\kappa),u=1,...,m\}$.
The uniqueness condition on $\mathcal G$ can be satisfied if there is a one-to-one correspondence (i.e., a bijection) between the elements of $\mathcal G$ and $\mathcal S_\mathcal G$.
In other words, the set $\mathcal G$ cannot include two or more waveforms with the same sign sequence.
The maximum size of $\mathcal G$ under the uniqueness condition will be investigated in Sec. IV.

\subsubsection{Paired Construction}
The waveforms in $\mathcal G$ can be included in pairs, i.e., if $g_u(t)\in \mathcal G$, then we also include $-g_u(t)$ in $\mathcal G$.
Although \emph{not} mandatory, this is helpful in system implementation and performance evaluation.
The pair $\{g_u(t), -g_u(t)\}$, referred to as an \emph{antipodal pair} hereinafter, can be obtained as (\ref{gs}) from the same $s(t)$ and a pair of antipodal amplitude-scaling factors.
The sign sequence of $-g_u(t)$ is $-\mathsf{sgn}(\mathbf{g}_{u}^\kappa)$.
In this setting we can divide $\mathcal G$ into two disjoint subsets with the same size as $\mathcal G_-=\left\{g_u(t),u=1,...,\frac{m}{2}\right\}$ and $\mathcal G_+=\left\{-g_u(t),u=1,...,\frac{m}{2}\right\}$, where $\mathcal G_+$ includes all waveforms in $\mathcal G$ satisfying $g_u(0^+)>0$.
Accordingly, all waveforms in $\mathcal G_-$ must satisfy $g_u(0^+)<0$.
Note that the paired construction must satisfy the uniqueness condition; otherwise, some antipodal pairs should be excluded.

\subsubsection{Restrictions on Zero-Crossings}

From (\ref{gs}), it is clear that, except for an amplitude-scaling and a time-scaling, a waveform $g_u(t)$ is determined by a realization of $S(t)$ and a parameter $k_0$ indicating the truncation interval.
The realization $s(t)$ is determined by the sequence $\{\tau_k\}$.
We impose a natural restriction on $\{\tau_k\}$ such that all $\tau_k$ satisfying $k<0$ and $k>\kappa-1$ (i.e., those outside the truncation interval) are equal to $k$, since varying them cannot introduce more waveforms with different zero-crossings.
Additionally, we can fix the truncation interval in our waveform construction by letting $k_0=0$ without loss of generality,
due to the fact that a time shifting of $\{\tau_k\}$ corresponds to a time shifting of $s(t)$ except for an amplitude-scaling and a possible sign-flipping.
This fact is formally expressed by Proposition 1 as follows.

\emph{Proposition 1: Let $\{\delta_k\}$, $k\in\mathbb{Z}$ be a fixed sequence satisfying $-\frac{1}{2}<\delta_k\leq\frac{1}{2}$.
Let $s(\boldsymbol{\delta},t)$ and $s\left(\boldsymbol{\delta}_{(i)},t\right)$ be two realizations of $S(t)$ determined by sequences $\{\tau_k=k+\delta_k\}$ and $\{\tau_k=k+\delta_{k+i}\}$, respectively, where $i$ is a finite integer. Then there exists a constant $c_i>0$ such that }
\begin{equation}
s\left(\boldsymbol{\delta}_{(i)},t-i\right)=(-1)^i c_i\cdot s(\boldsymbol{\delta},t), \mspace{4mu} \forall t\in\mathbb R.
\end{equation}
\begin{IEEEproof}
The proof is given in Appendix A.
\end{IEEEproof}

\subsection{Power Spectral Density and Soft Truncation}

Since truncation always results in spectrum broadening, we must calculate the fractional power containment bandwidth $W_\eta$ of $X(t)$ to evaluate the spectral efficiency achieved for a given $\eta$.
In particular, when $\mathcal G$ consists of antipodal pairs, we have the following result.

\emph{Proposition 2}: If the set $\mathcal G$ consists of $\frac{m}{2}$ antipodal pairs of waveforms and the input symbol $U$ is uniformly distributed over $\mathcal U=\{1,...,m\}$, the PSD of $X(t)$ is given by
\begin{equation} \label{eq_pf}
\mathsf{S}(f)=\frac{1}{m\kappa T_{\mathrm N}} \sum_{u=1}^{m} \left|\hat{g}_u(f)\right|^2,
\end{equation}
and the power of $X(t)$ satisfies
\begin{equation}
\label{PE}
\mathsf P=\frac{1}{m\kappa T_\mathrm N}\sum\limits_{u=1}^m \int_{-\infty}^{\infty}g_u^2(t)\mathrm dt.
\end{equation}
\begin{IEEEproof}
The proof is given in Appendix B.
\end{IEEEproof}

According to Proposition 2, the power and the PSD of $X(t)$ is determined by $\mathcal G$.
As the truncation length $\kappa T_{\mathrm{N}}$ increases, the out-of-band power of $X(t)$ with respect to $[-W_\mathrm{N},W_\mathrm{N}]$ decreases, which helps to improve the spectral efficiency.
However, increasing the truncation length causes a large detection complexity.
For practical truncation lengths, to control the out-of-band emission of $X(t)$ with respect to the nominal bandwidth, we may perform \emph{soft truncation}; that is, we introduce a window function $h(t)$ which satisfies $0\leq h(t)\leq 1$ in $(0,\kappa T_{\mathrm N}]$ into the truncation as
\begin{equation}
\label{gss}
g_u(t)=\varphi_u s\left(\frac{t+k_0-\frac{1}{2}}{T_{\mathrm N}}\right)\cdot\mathds{1}_{\left(0,\kappa T_\mathrm N\right]}(t)\cdot h(t).
\end{equation}
The soft truncation does not change the sign of $s(t)$ for $t\in\mathbb R$.
It affects the achievable information rate by changing the transition probabilities of the resulting DMC (\ref{DMC}).
It also affects the spectral efficiency by changing the PSD.
Finding the optimal window function for a given time duration is therefore difficult.
Here, we employ a raised cosine window \cite{Lapidoth2017} as
\begin{figure*}
\begin{equation} \label{eq_second}
h_{\mathrm{RC}}(t)=\left\{
\begin{aligned}
&1,    &　&{0\leq|t|\leq \frac{(1-\alpha)T}{2}}\\
&\frac{1}{2}\left(1+{\rm cos}\left(\frac{2\pi}{\alpha T}\left( |t|-\frac{(1-\alpha)T)}{2}\right) \right)\right),   &　&{\frac{(1-\alpha)T}{2}<|t|\leq\frac{T}{2}}\\
&0,                              &　&{|t|>\frac{T}{2}},
\end{aligned} \right.
\end{equation}
\end{figure*}
where $0\leq\alpha\leq 1$ is the roll-off factor.
Based on (\ref{eq_second}), the window function in (\ref{gss}) is given by scaling and shifting in time as $h(t)=h_{\mathrm{RC}}\left(\frac{T}{\kappa T_{\mathrm{N}}}t-\frac{T}{2}\right)$.
Fig. \ref{waveform} and Fig. \ref{set_PSD} are examples illustrating the effect of soft truncation, where we let $\kappa=3$, $n=3$, and construct waveforms using the uniform zero-crossing pattern.
In Fig. \ref{waveform}, we show different versions of waveforms obtained by truncating the same $s(t)$ with raised cosine windows of different roll-off factors.
Fig. \ref{set_PSD} shows how the soft truncation changes the PSD of $X(t)$
where the waveform sets used are obtained following the same procedure except the roll-off factors in soft truncation.
In fact, the PSD is scaled by a factor of $|\hat{h}(f)|^2$ according to Proposition 2.
\begin{figure}
\includegraphics[width=1\columnwidth]{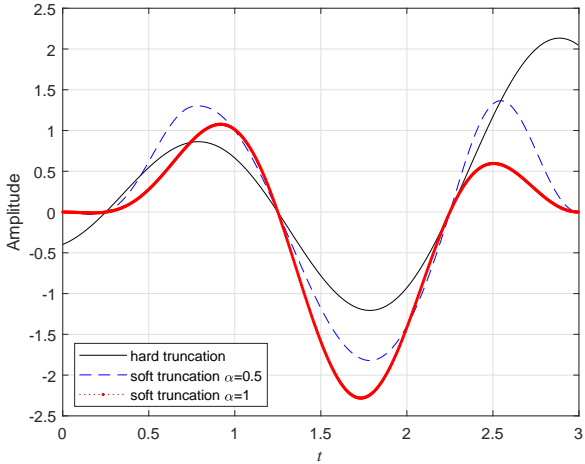}
\caption{Examples of $g_u(t)$ under soft truncation of different roll-off factors.}
\label{waveform}
\end{figure}
\begin{figure}
\includegraphics[width=1\columnwidth]{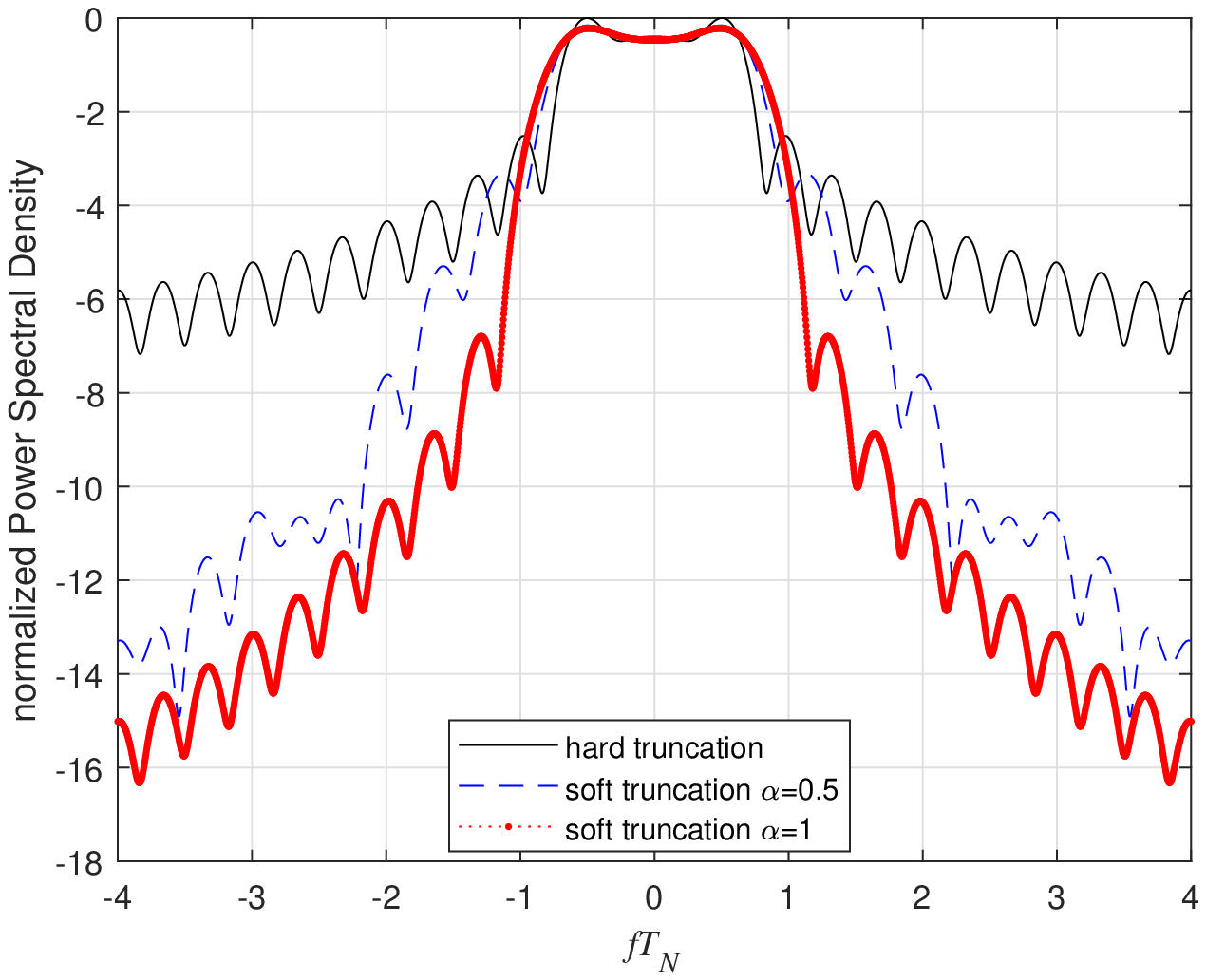}
\caption{Examples of $\mathsf S(f)$ under soft truncation of different roll-off factors.}
\label{set_PSD}
\end{figure}

\subsection{A Summary of Waveform Set Construction}
We have introduced several techniques in the construction of the waveform set $\mathcal G$.
We summarize these techniques in Table \uppercase\expandafter{\romannumeral1} to clarify their methods, main parameters, and roles in our problem.

\begin{table*}[tbp]
\centering
\renewcommand{\cellset}
{\renewcommand{\arraystretch}{1}}
\centering
\caption{Waveform Set Construction: A Summary}
\scalebox{1}{
\begin{tabular}{l|l|l|l}
\hline
TECHNIQUE &METHOD &PARAMETER&ROLE\\\hline
&&\vspace{-1.0em}\\\hline
Truncation              &Adjusting length of $g_u(t)$           &$\kappa$ (Nyquist intervals) &Simplifying transmitter  \\\hline
Soft truncation         &Truncating by window functions with roll-off  &$\alpha$   &Reducing bandwidth occupied       \\\hline
Zero-crossing design    &\makecell[l]{Adjusting number of possible zero-crossing positions \\Designing uniform or nonuniform patterns (\ref{tl})/(\ref{tau_Unequally})}       &\makecell[l]{$n$ (per Nyquist interval)\\ $\{\Delta[l]\}$}&Boosting information rate \\\hline
Paired construction     &Including waveforms in antipodal pairs        &N/A        &Boosting information rate         \\\hline
Amplitude-scaling       &Adjusting energy of $g_u(t)$                &$\varphi_u$   &Energy normalization\\\hline
Time-scaling            &Adjusting bandwidth of $g_u(t)$             &$T_\mathrm N$ &Matching available bandwidth \\\hline
\end{tabular}}
\end{table*}

\section{Information Rate: Asymptotic Analysis at High SNR}

The performance of the proposed transceiver is determined by the waveform set ${\mathcal G}$ and the input distribution $p(u)$.
This section focuses on the information rate in the high-SNR regime.

\emph{Lemma 1}: The high-SNR limit of the information rate $R$ given in (\ref{IR}) is
\begin{equation}
\lim\limits_{\mathsf{SNR}\to\infty}R=\frac{1}{\kappa T_\mathrm N}H(U).
\end{equation}

\begin{IEEEproof}
By noting that the absolute values of the arguments in the Q functions in (\ref{pyx}) increase continuously and monotonically with $\mathsf {SNR}$ for all $1\leq k\leq \kappa$ and $1\leq \ell\leq n$, we obtain
\begin{equation}
\lim\limits_{\mathsf{SNR}\to\infty} p(\mathbf b^\kappa|u)=
\left\{
\begin{aligned}
&1,& &\mathbf b^\kappa=\mathsf{sgn}(\mathbf g_{u,k})\\
&0, & &\textrm{otherwise},
\end{aligned}\right.
\end{equation}
and
\begin{align}\label{pub}
\lim\limits_{\mathsf{SNR}\to\infty} p(u,\mathbf b^\kappa)=&\lim\limits_{\mathsf{SNR}\to\infty} p(\mathbf b^\kappa|u)p(u)\notag\\
=&
\left\{
\begin{aligned}
&p(u),& &\mathbf b^\kappa=\mathsf{sgn}(\mathbf g_{u,k})\\
&0, & &\textrm{otherwise}.
\end{aligned}\right.
\end{align}
Under the uniqueness condition, we have $\mathsf{sgn}(\mathbf g_{u,k})\neq \mathsf{sgn}(\mathbf g_{u',k})$ if $u\neq u'$.
So
\begin{align}\label{pb}
\lim\limits_{\mathsf{SNR}\to\infty} p(\mathbf b^\kappa)&=\lim\limits_{\mathsf{SNR}\to\infty} \sum\limits_{u'=1}^m p(u',\mathbf b^\kappa)\notag\\
&=
\left\{
\begin{aligned}
&p(u),& &\mathbf b^\kappa=\mathsf{sgn}(\mathbf g_{u,k})\\
&0, & &\textrm{otherwise}.
\end{aligned}\right.
\end{align}
From (\ref{pub}) and (\ref{pb}), the high-SNR limits of $H(U,\mathbf{B}^{\kappa})$ and $H(\mathbf{B}^{\kappa})$ are both $H(U)$.
Noting that
\begin{equation}
I(U;\mathbf{B}^{\kappa})=H(U)+H(\mathbf{B}^{\kappa})-H(U,\mathbf{B}^{\kappa}),
\end{equation}
we complete the proof.
\end{IEEEproof}

Since $H(U)\leq \log m$ with equality if $U$ is uniformly distributed, the high-SNR information rate is dominated by the size of $\mathcal G$.
For given $\kappa$ and $n$, we use $M(\kappa,n)$ to denote the maximum size of $\mathcal G$,
i.e., the maximum number of waveforms that can be constructed from truncating the realizations of $S(t)$ subject to the uniqueness condition. From Lemma 1, we have that information rate
\begin{equation}
\label{AR}
R=\frac{\log M(\kappa,n)}{\kappa T_\mathrm{N}}
\end{equation}
is asymptotically achievable at high SNR by employing a set $\mathcal G$ with size $|\mathcal G|=M(\kappa,n)$ and letting $U$ be uniformly distributed over $\{1,...,M(\kappa,n)\}$, and that this asymptotic information rate cannot be improved with the transceiver structure described in Sec. II.

Now the problem boils down to determining the maximum size $M(\kappa,n)$ for $\mathcal G$ and providing an explicit design to achieve it.
To determine $M(\kappa,n)$, recalling that $|\mathcal S_\mathcal G|=|\mathcal G|=|\mathcal U|=m$, we consider the maximum size of $\mathcal S_{\mathcal G}$.
From the facts that in each Nyquist interval of $S(t)$ there is only one zero-crossing, and that zero-crossing does not necessarily cause a sign flip (this is due to the integrate-and-dump filter used),
we have a simple observation: \emph{in each $n$-tuple} $\mathsf{sgn}(\mathbf{g}_{u,k})$ \emph{in a sign sequence} $\mathsf{sgn}(\mathbf{g}_{u}^\kappa)$, \emph{the sign flips at most once}
(here, a flip from $\mathrm{sgn}(g_{u,k}[n])$ to $\mathrm{sgn}(g_{u,k+1}[1])$ is counted as a flip in the $n$-tuple $\mathsf{sgn}(\mathbf{g}_{u,k})$).
We call this observation an \emph{admissibility condition} for $\mathcal S_{\mathcal G}$, which should be satisfied if a sequence ${\mathbf b}^\kappa$ drawn from all $2^{\kappa n}$ sequences in the output alphabet is included in ${\mathcal S}_{\mathcal G}$.
Take $n=4$ for example. When $\kappa=1$, $[+ + - -]$ is admissible but $[+ + - +]$ is not because it has two sign flips; when $\kappa=2$, ${[+ + - - , - + + +]}$, ${[- - - -, - - + +]}$, and ${[+ + + + , + + + +]}$ are all admissible (where we use comma to divide successive $n$-tuples), but ${[+ + - -, + + + -]}$ is not.
In general, we will prove that
\begin{equation}
\label{M}
M(\kappa,n)=2(n+1)^{\kappa}-2^\kappa.
\end{equation}
Combining this with (\ref{AR}), we obtain the main result of this section as follows, which holds for arbitrary zero-crossing patterns employed in the construction of $X(t)$.

\emph{Proposition 3: The high-SNR limit of the maximum information rate of the transceiver structure described in Sec. II is}
\begin{align}
\label{Rmax}
R_\mathrm{max}=\frac{1}{T_\mathrm{N} }\left(\log_2\left((n+1)^\kappa-2^{\kappa-1}\right)^{\frac{1}{\kappa}}+\frac{1}{\kappa}\right) \mspace{4mu} \textrm{bits/s}.
\end{align}

\begin{IEEEproof} In light of Lemma 1 and (\ref{AR}), we only need a proof of (\ref{M}), which is obtained by combining 1) and 2) as follows.

1) \emph{Upper bound}:
Based on the admissibility condition for $\mathcal S_{\mathcal G}$, we prove $M(\kappa,n)\leq2(n+1)^{\kappa}-2^\kappa$ as follows.
According to the definition of $S(t)$, a waveform obtained from (\ref{gs}) satisfies either $g_u(0^+)>0$ or $g_u(0^+)<0$.
We first consider those satisfying $g_u(0^+)>0$.
From the property of $S(t)$ given by (\ref{sign}), we can infer that $g_u(((k-1)T_{\mathrm N})^+)$ is positive for odd $k$ and negative for even $k$.
Hence, for odd $k$, a flip in $\mathsf{sgn}(\mathbf g_{u,k})$, if exists, must be from $+$ to $-$, no matter which zero-crossing pattern is employed.
The case of even $k$ is similar except that a flip must be from $-$ to $+$.
There are at most $n+1$ possible choices for $\mathsf{sgn}(\mathbf g_{u,k})$, including $n-1$ tuples with only one sign flip, the all-one tuple, and the opposite of it.\footnote{An example for $n=4$ and even $k$ is given in (\ref{sskn}). In general, there exist $2n$ length-$n$ tuples with at most one sign flip, including $n-1$ ones with a flip from $+$ to $-$, $n-1$ ones with a flip from $+$ to $-$, the all-one tuple and its opposite.}
Therefore, $\mathcal S_{\mathcal G}$ contains at most $(n+1)^{\kappa}$ sign sequences which correspond to waveforms satisfying $g_u(0^+)>0$.
We denote the set of these sequences by $\mathcal S_+$.
By symmetry, $\mathcal S_{\mathcal G}$ contains at most $(n+1)^{\kappa}$ sign sequences that correspond to waveforms satisfying $g_u(0^+)<0$, and we denote the set of these sequences by $\mathcal S_-$.
The union $\mathcal S_+\cup\mathcal S_-$ includes all admissible sign sequences.
However, the sign sequences that satisfy $\mathsf{sgn}(\mathbf{g}_{u,k}) \in \{\mathbf{1},-\mathbf{1}\}$, $k=1,...,\kappa$ are included in both $\mathcal S_+$ and $\mathcal S_-$, and there are $2^\kappa$ such sequences.
Therefore, the maximum size of $\mathcal S_\mathcal G$, namely $M(\kappa,n)$, is upper bounded by $|\mathcal S_+\cup\mathcal S_-|=|\mathcal S_+|+|\mathcal S_-|-|\mathcal S_+\cap\mathcal S_-|$, which is at most $2(n+1)^{\kappa}-2^\kappa$.

2) \emph{Achievability}:
Based on the nonuniform zero-crossing pattern given in Sec. III-B,
we exhibit an explicit paired design of $\mathcal G$ satisfying $|\mathcal G|=2(n+1)^{\kappa}-2^\kappa$ and the uniqueness condition.
We truncate realizations of $S(t)$ as (\ref{gs}) over the interval $\big(-\frac{1}{2},\kappa-\frac{1}{2}\big]$, where
the nonuniform zero-crossing pattern (\ref{tau_Unequally})
is employed.
This yields a set $\mathcal G_-$ including $(n+1)^{\kappa}$ waveforms satisfying $g_u(0^+)<0$.
Consider a new set of $(n+1)^{\kappa}$ waveforms denoted by $\mathcal G_+$, which consists of the opposites of all waveforms in $\mathcal G_-$ (i.e., if $g_u(t)\in \mathcal G_-$, then $-g_u(t)\in{\mathcal G}_+$).
From the property of the nonuniform zero-crossing pattern discussed in Sec. III-B, and the property of $S(t)$ given by (\ref{sign}),
the set of the sign sequences of $\mathcal G_-$ and $\mathcal G_+$ is exactly $\mathcal S_-$ and $\mathcal S_+$, respectively.\footnote{We may also employ soft truncation as (\ref{gss}) in the proof herein, and the property of the nonuniform zero-crossing pattern can still be utilized by letting $\lambda$ be sufficiently small. It can further be shown that for any finite truncation length $\kappa$, there exists a strictly positive $\lambda$ to satisfy the nonuniform zero-crossing pattern.}
Take a sign sequence $\mathbf {b}^\kappa\in \mathcal S_-$ for example, the corresponding waveform in $\mathcal G_{-}$ is the one obtained by i) letting
\begin{equation}
\tau_k=\left\{
\begin{aligned}
&k-\frac{1}{2}+\Delta[l],& &k=0,...,\kappa-1\\
&k, & &\textrm{otherwise},
\end{aligned}\right.
\end{equation}
in $S(t)$, where $\Delta[l]$ is given by (\ref{tau_Unequally}) and
\begin{equation}
l=\frac{1}{2}\left(n+(-1)^{k}\sum\limits_{\ell=1}^n b_k[\ell]\right),
\end{equation}
ii) truncating the obtained realization $s(t)$ over the interval $\big(-\frac{1}{2},\kappa-\frac{1}{2}\big]$, and iii) shifting and scaling it in time.
Clearly, $\mathcal G_-\cap {\mathcal G}_+=\emptyset$. So we obtain a set $\mathcal G_-\cup {\mathcal G}_+$ with the size of $2(n+1)^{\kappa}$.
Since there are $2^{\kappa}$ sequences belonging to both $\mathcal S_-$ and $\mathcal S_+$, to comply with the uniqueness condition, we correspondingly remove $2^{{\kappa}-1}$ waveforms satisfying $\mathsf{sgn}(\mathbf{g}_{u,k}) \in \{\mathbf{1},-\mathbf{1}\}$ in $\mathcal G_-$ and their opposites in ${\mathcal G}_+$.
The set of the sign sequences of the removed waveforms is exactly $\mathcal S_-\cap\mathcal S_+$, and the remaining waveforms form $(n+1)^{\kappa}-2^{\kappa-1}$ antipodal pairs, which constitute the set $\mathcal G$ we need. Therefore the size $|\mathcal G|=|\mathcal S_-\cup\mathcal S_+|=2(n+1)^{\kappa}-2^\kappa$ can be achieved without violating the uniqueness condition.
\end{IEEEproof}

\emph{Remark 1}:
From Proposition 3, a rate $\frac{1}{T_\mathrm{N} }\log_2(n+1)$ bits/s can be achieved in the limit of large $\kappa$.
Since the difference between the spectra of $X(t)$ and $S(t)$ also vanishes as the truncation length $\kappa T_{\mathrm N}$ grows without bound, we can infer that the limit of the maximum spectral efficiency achieved by our transceiver is $2\log_2(n+1)$ bits/s/Hz.
This is consistent with the result in \cite{Shamai1994} that $\log_2(n+1)$ bits per Nyquist interval can be achieved in the noiseless case.

When the uniform zero-crossing pattern (\ref{tl}) is employed, a set $\mathcal G^\mathrm{unif}_{-}$ including $n^{\kappa}$ waveforms satisfying $g_u(0^+)<0$ can be obtained by truncating $S(t)$ over the interval $\big(-\frac{1}{2},\kappa-\frac{1}{2}\big]$.
We use $\overline{\mathcal G}^\mathrm{unif}_{+}$ to denote the set which consists of the opposites of all waveforms in $\mathcal G^\mathrm{unif}_{-}$.
It can be shown that both sets satisfy the uniqueness condition, and that the sets of the sign sequences of them are disjoint: $\mathcal S^\mathrm{unif}_{-}\cap \overline{\mathcal S}^\mathrm{unif}_{+}=\emptyset$.
Thus, based on the uniform zero-crossing pattern, we can construct a waveform set $\mathcal G=\mathcal G^\mathrm{unif}_{-}\cup\overline{\mathcal G}^\mathrm{unif}_{+}$ satisfying $|\mathcal G|=2n^{\kappa}$,
and asymptotically achieve
\begin{align}
R^\mathrm{unif}=\frac{1}{T_\mathrm{N}}\left(\log_2 n+\frac{1}{\kappa}\right) \textrm{bit/s},
\end{align}
which is smaller than $R_\mathrm{max}$ given in (\ref{Rmax}) when $n>1$ and $\kappa>1$.
For large $\kappa$ we have $R^\mathrm{unif}\approx \frac{1}{T_\mathrm{N} }\log_2 n$.

\section{Performance Evaluation: Numerical Results}\label{subsec_achieve_rates}

\subsection{Spectral Efficiency: Finite SNR Performance}\label{SEFinite}

Although the high-SNR performance of our transceiver is dominated by the size of $\mathcal G$,
when we focus on a given SNR,
several other aspects should be taken into consideration, including the zero-crossing pattern,
the roll off factor $\alpha$, the input distribution $p(u)$, and the amplitude-scaling factors $\{\varphi_u\}$.
However, jointly optimizing these parameters with the goal of maximizing spectral efficiency appears intractable and non-intuitive.
In this section we give some performance evaluation examples for our transceiver numerically under the following simplifications.
First, besides constructing the waveforms in $\mathcal G$ in pairs, we further adjust $\varphi_u$ to normalize the waveforms as
$\int_{-\infty}^{\infty}g_u^2(t){\mathrm d}t=\mathcal E,\mspace{4mu}u=1,...,m$.
Second, we let $U$ be uniformly distributed, which is asymptotically optimal at high SNR but suboptimal in general (see Remark 2 for more discussions).
These settings also simplify PSD calculations based on Proposition 2.
Additionally, when employing the nonuniform zero-crossing pattern given in (\ref{tau_Unequally}) we always let $\lambda=1/4$, which may be further optimized in future studies.

For a given waveform set $\mathcal G$, we can evaluate the achievable spectral efficiency $\mathsf {SE}=\frac{R}{W_\eta}$ where $R$ and $W_\eta$ are determined by (\ref{IR}) and (\ref{Weta}), respectively.
We set $\eta=0.95$ in all our numerical experiments.
For given $\kappa$, $n$, $\alpha$, and a zero-crossing pattern,
we can let $\mathcal G$ contain as many waveforms as possible under the uniqueness condition.
However, it is possible to improve the spectral efficiency by including only a part of the available waveforms into $\mathcal G$,
because the bandwidth $W_\eta$ can be reduced if we exclude some waveforms with relatively worse concentration in frequency domain.
As an example, we use the following heuristic procedure to optimize the spectral efficiency for a given SNR, where $m$ is even satisfying $2\leq m \leq M$:
\begin{enumerate}[i)]
  \item For a given bandwidth $W$ (typically greater than $W_{\mathrm N}$), calculate in-band energies for all waveform pairs, and sort the pairs in descending order of in-band energy.
  \item Let $\mathcal G$ be the set of the first $m/2$ pairs, calculate the corresponding PSD of $X(t)$ according to Proposition 2, and calculate $\eta$ with respect to $W$.
  \item Adjust $W$ (by, e.g., a bisection search) and repeat i) and ii) until $\eta$ is slightly larger than the target 0.95 (note that the sorting result may vary if $W$ changes);
  \item Calculate the spectral efficiency achieved according to (\ref{IR}) and (\ref{Weta}).
\end{enumerate}
Numerical results on the achievable spectral efficiency for different values of $\alpha$ and $m$ are shown in Fig. \ref{fig_23_dB}. The simulation parameters are shown in Table \uppercase\expandafter{\romannumeral2}.
In this example we let $\kappa=3$, $n=4$, $\mathsf{SNR}=25$ dB, and use the uniform zero-crossing pattern so that we have at most $M=2n^\kappa=128$ waveforms.
It is shown that when $\alpha$ is relatively small (e.g., $0.3$ or less), using part of the available waveforms can be good enough, and using all available waveforms is indeed detrimental.
Specifically, when $\alpha=0$ or $\alpha=0.1$, using about a half of the available waveforms can outperform all other settings, and a spectral efficiency of roughly 1.4 bits per dimension can be achieved.

\emph{Remark 2}:
{
\begin{table*}[tbp]
\renewcommand{\cellset}
{\renewcommand{\arraystretch}{1}}
\centering
\caption{A Summary of Simulation Parameters}
\scalebox{1}{
\begin{tabular}{l|l|l|l|l|l}
\hline
Figures                     &Zero-Crossing Pattern  &$\kappa$   &$n$              &Parameters to be Optimized &Goal of Optimization                            \\\hline
&&&&&\vspace{-1.0em}\\\hline
\ref{fig_23_dB}                  &Uniform                &$3$        &$4$              &\multirow{5}{*}{\makecell[cl]{$\alpha\in\{0,0.1,...,1\}$\\$M/2\in\{1,...,64\}$}}&\multirow{5}{*}{\makecell[cl]{Spectral efficiency\\ $\mathsf {SE}=\frac{R}{W_\eta}$}}   \\\cline{1-4}
\ref{fig_dif_Ny_c_cap}           &Uniform                &$1, 2, 3$  &$4$              & &                                        \\\cline{1-4}
\ref{fig_dif_Ny_c_cap1}          &Nonuniform              &$1, 2, 3$  &$4$              & &                                        \\\cline{1-4}
\ref{fig_dif_overs_c_cap}        &Uniform                &$3$        &$2, 3, 4, 5, 6$  & &                                        \\\cline{1-4}
\ref{fig_dif_overs_c_cap1}       &Nonuniform              &$3$        &$1, 2, 3, 4, 5$  & &                                        \\\hline
\end{tabular}}
\end{table*}
}
We use the above heuristic procedure because finding the optimal solution that maximizes the spectral efficiency is extremely difficult.
If $m/2$ pairs of waveforms in all $M/2$ available ones are included in $\mathcal G$, we need to exhaustively calculate $\binom{M/2}{m/2}$ possible values of spectral efficiency for comparison, a computationally infeasible task when $m$ is large.
Fortunately, the loss of our heuristic procedure is usually limited.
As an example, Fig. \ref{BA} provides some numerical results where we use the same setting as Fig. \ref{fig_23_dB} but consider only the case $\alpha=0$.
For each SNR we provide 100 examples of the information rate $R$ given by (\ref{IR}), where we let $p(u)=\frac{1}{m}, \forall u\in\mathcal U$, and the transition probability $p(\mathbf b^\kappa|u)$ for each example is determined by a set $\mathcal G$ consisting of $32$ pairs \emph{randomly} chosen from all $64$ available ones.
For comparison, using the Blahut-Arimoto algorithm \cite{CT06}, we also calculate the capacity of the corresponding DMC (\ref{DMC}) when the set $\mathcal G$ contains all $M/2=n^\kappa=64$ pairs of waveforms.
The results show that using a part of available waveforms with a uniform input distribution can be good enough at low to moderate SNR, and the performance differences among different randomly generated sets $\mathcal G$ are usually small.
Nevertheless, as SNR increases, we should correspondingly increase $|\mathcal G|$ to improve the performance.
\begin{figure}
\includegraphics[width=1\columnwidth]{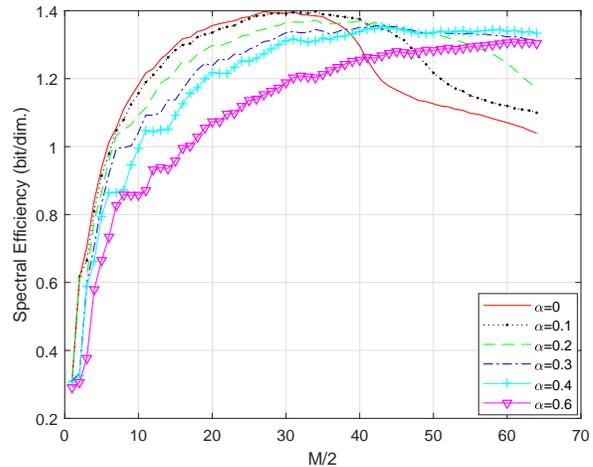}
\caption{Achievable {spectral efficiency} at $\mathsf{SNR}=25$ dB, $\kappa=3$, $n=4$, under uniform zero-crossing pattern.}
\label{fig_23_dB}
\end{figure}
\begin{figure}
\includegraphics[width=1\columnwidth]{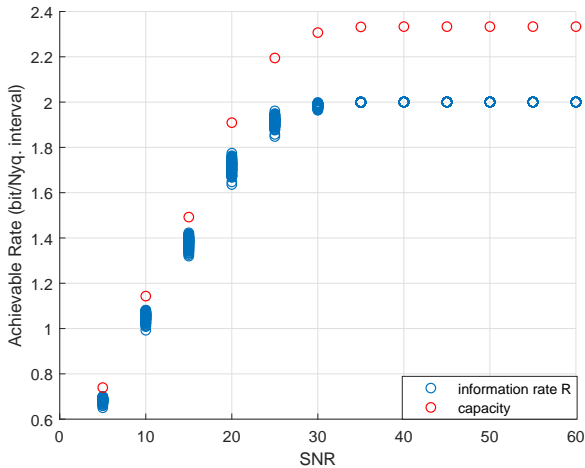}
\caption{{Performance analysis based on the DMC: i) capacity of the DMC, $m=M=128$, where $M$ is the maximum size of $|\mathcal G|$}; ii) information rates achieved by 64 randomly chosen inputs with equal probability. }
\label{BA}
\end{figure}

\emph{Remark 3}:
We may include waveforms with the same sign sequences (e.g. those obtained when employing the nonuniform zero-crossing pattern) into the proposed heuristic procedure by a small variation: In step ii), we should choose the first $m/2$ pairs that satisfy the uniqueness condition.
This may slightly improve the achievable spectral efficiency.

\begin{figure*}
\begin{minipage}[t]{0.48\linewidth}
\centering
\includegraphics[width=1\columnwidth]{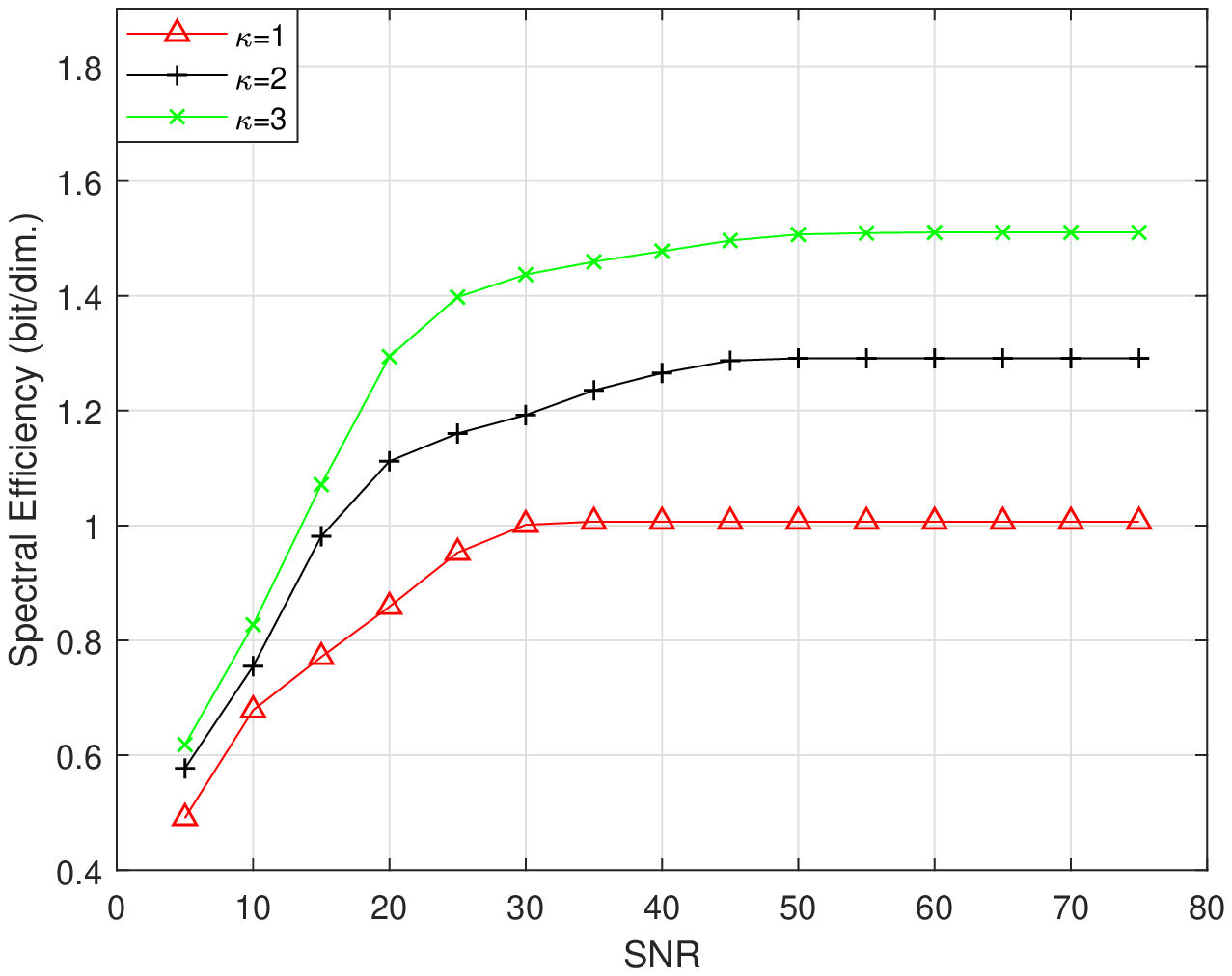}
\caption{Achievable { spectral efficiency} versus $\mathsf {SNR}$, $n=4$, uniform zero-crossing pattern.}
\label{fig_dif_Ny_c_cap}
\end{minipage}\hspace{5mm}
\begin{minipage}[t]{0.48\linewidth}
\centering
\includegraphics[width=1\columnwidth]{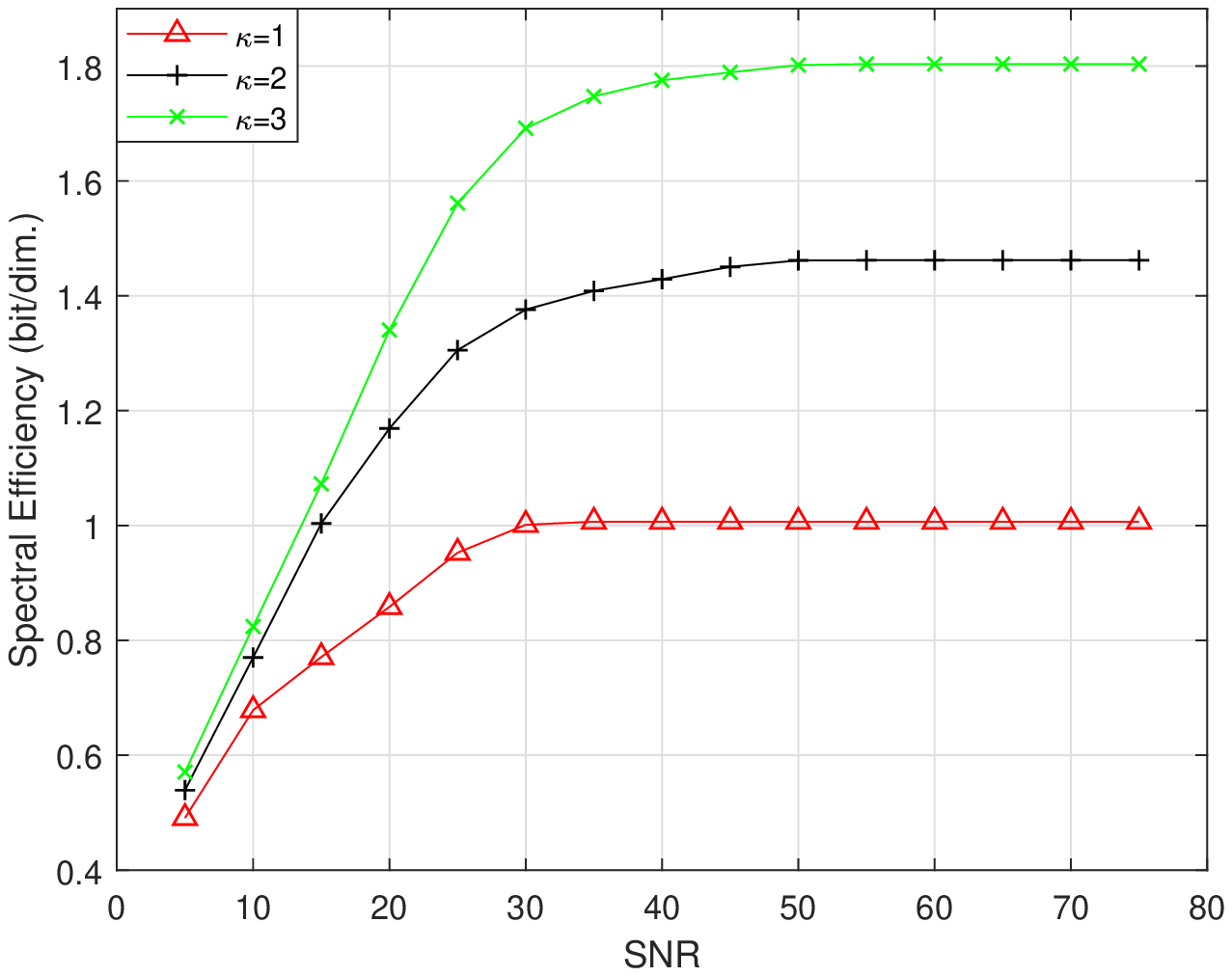}
\caption{Achievable { spectral efficiency} versus $\mathsf {SNR}$, $n=4$, nonuniform zero-crossing pattern.}
\label{fig_dif_Ny_c_cap1}
\end{minipage}
\end{figure*}
\begin{figure*}
\begin{minipage}[t]{0.48\linewidth}
\centering
\includegraphics[width=1\columnwidth]{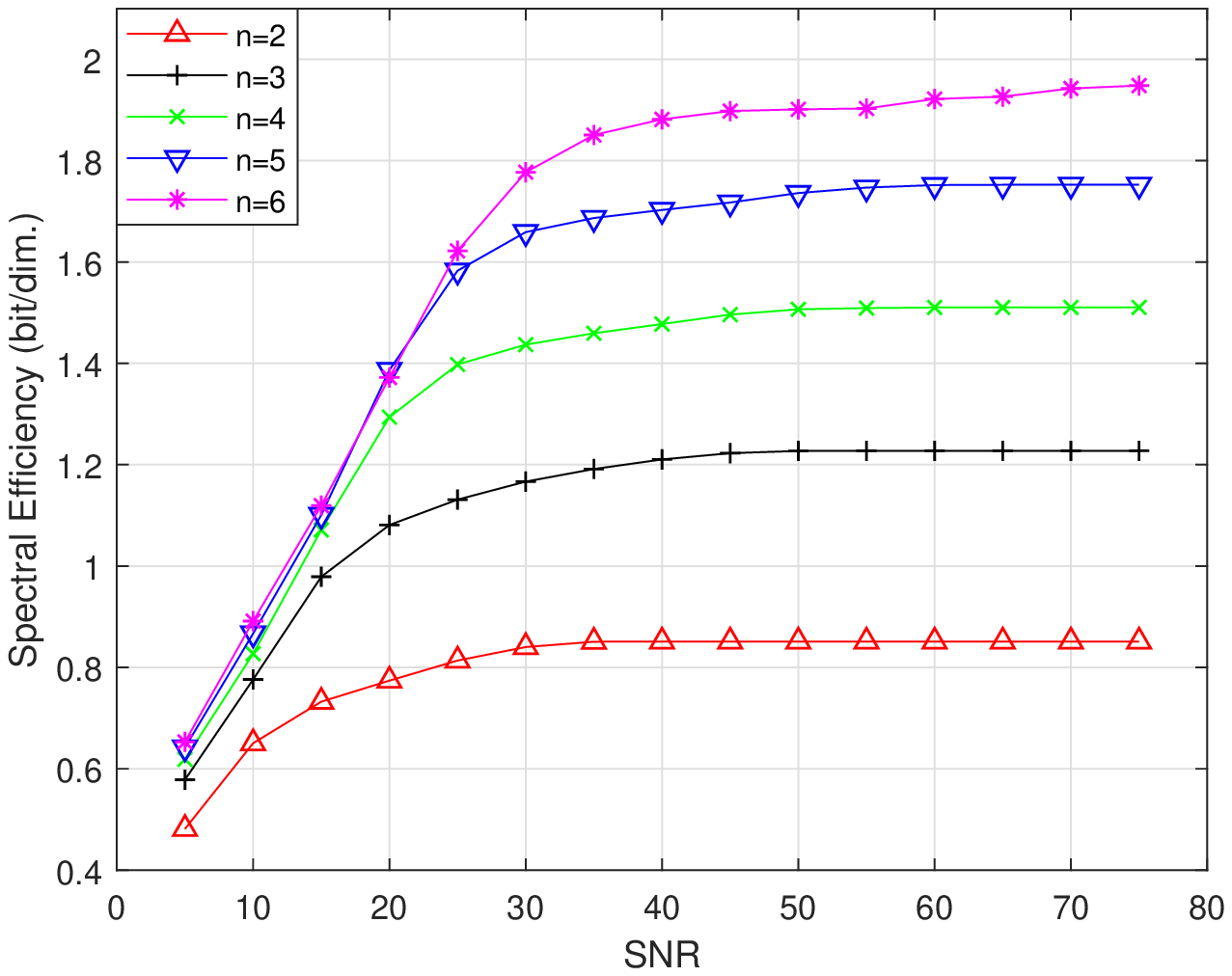}
\caption{Achievable { spectral efficiency} versus $\mathsf {SNR}$, $\kappa=3$, uniform zero-crossing pattern.}
\label{fig_dif_overs_c_cap}
\end{minipage}\hspace{5mm}
\begin{minipage}[t]{0.48\linewidth}
\centering
\includegraphics[width=1\columnwidth]{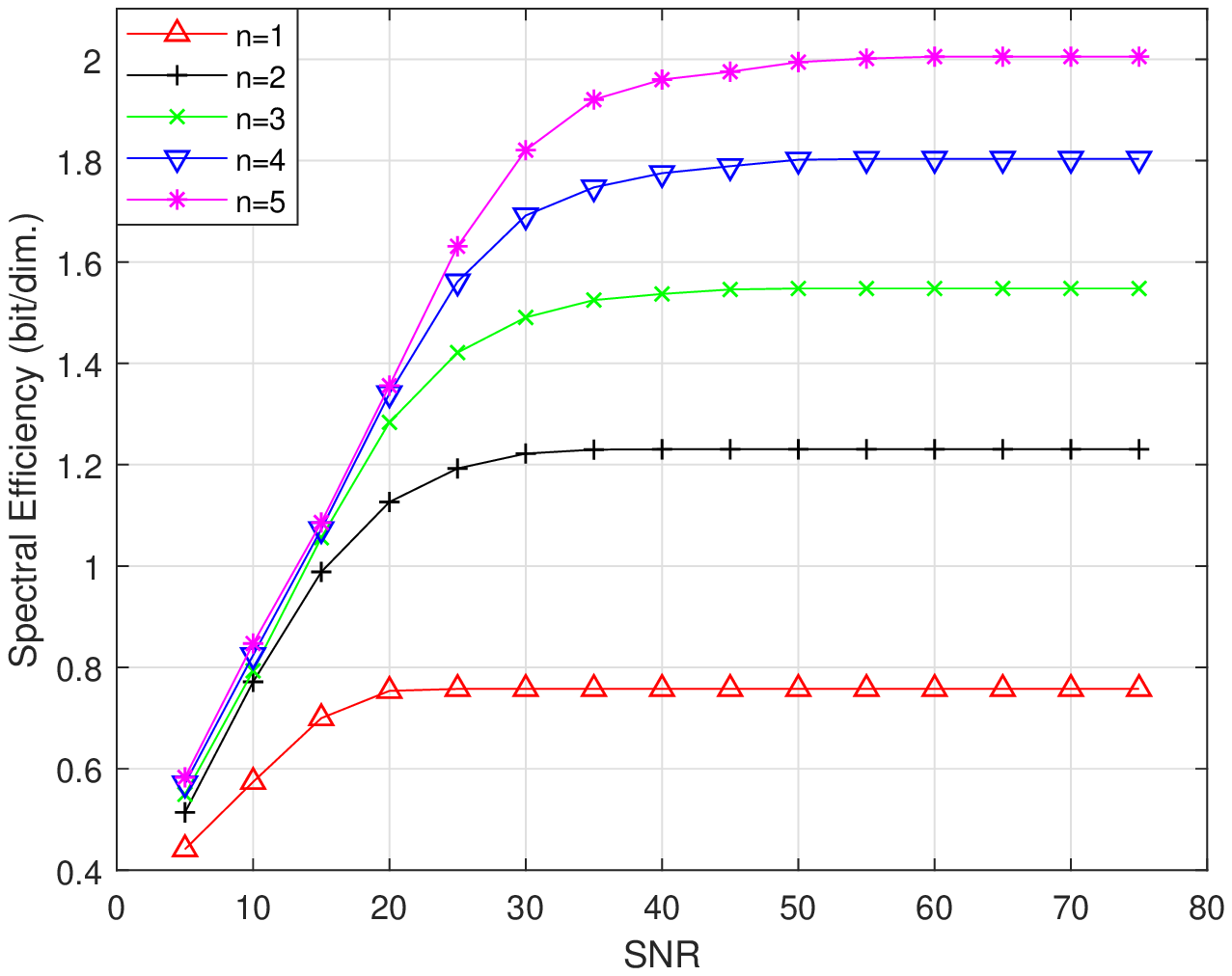}
\caption{Achievable { spectral efficiency} versus $\mathsf {SNR}$, $\kappa=3$, nonuniform zero-crossing pattern.}
\label{fig_dif_overs_c_cap1}
\end{minipage}
\end{figure*}

In Fig. \ref{fig_dif_Ny_c_cap}-\ref{fig_dif_overs_c_cap1}, we provide numerical results of the achieved spectral efficiency optimized by the proposed heuristic procedure with simulation parameters in Table \uppercase\expandafter{\romannumeral2}.
It is shown that considerable performance improvement can be obtained by increasing the oversampling factor.
For higher oversampling factors, the spectral efficiency reaches saturation at higher SNR.
Moreover, increasing the truncation length improves the spectral efficiency at the cost of increased transceiver complexity.

\subsection{Spectral Efficiency: High-SNR Limit}
\begin{figure*}\centering
\subfigure { \label{capacity_095}
\includegraphics[width=6.5in]{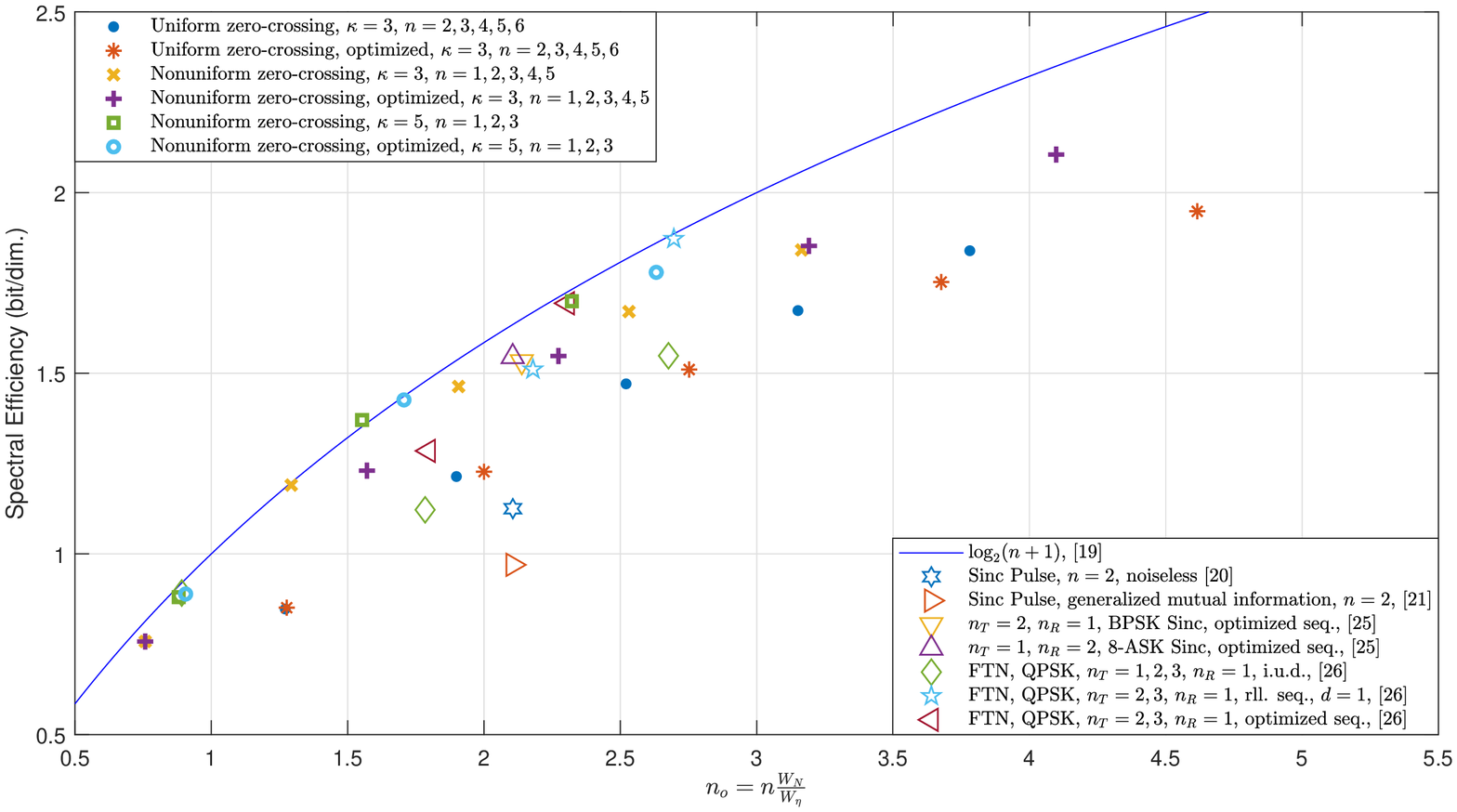}
}
\subfigure {\label{capacity_090}
\includegraphics[width=6.5in]{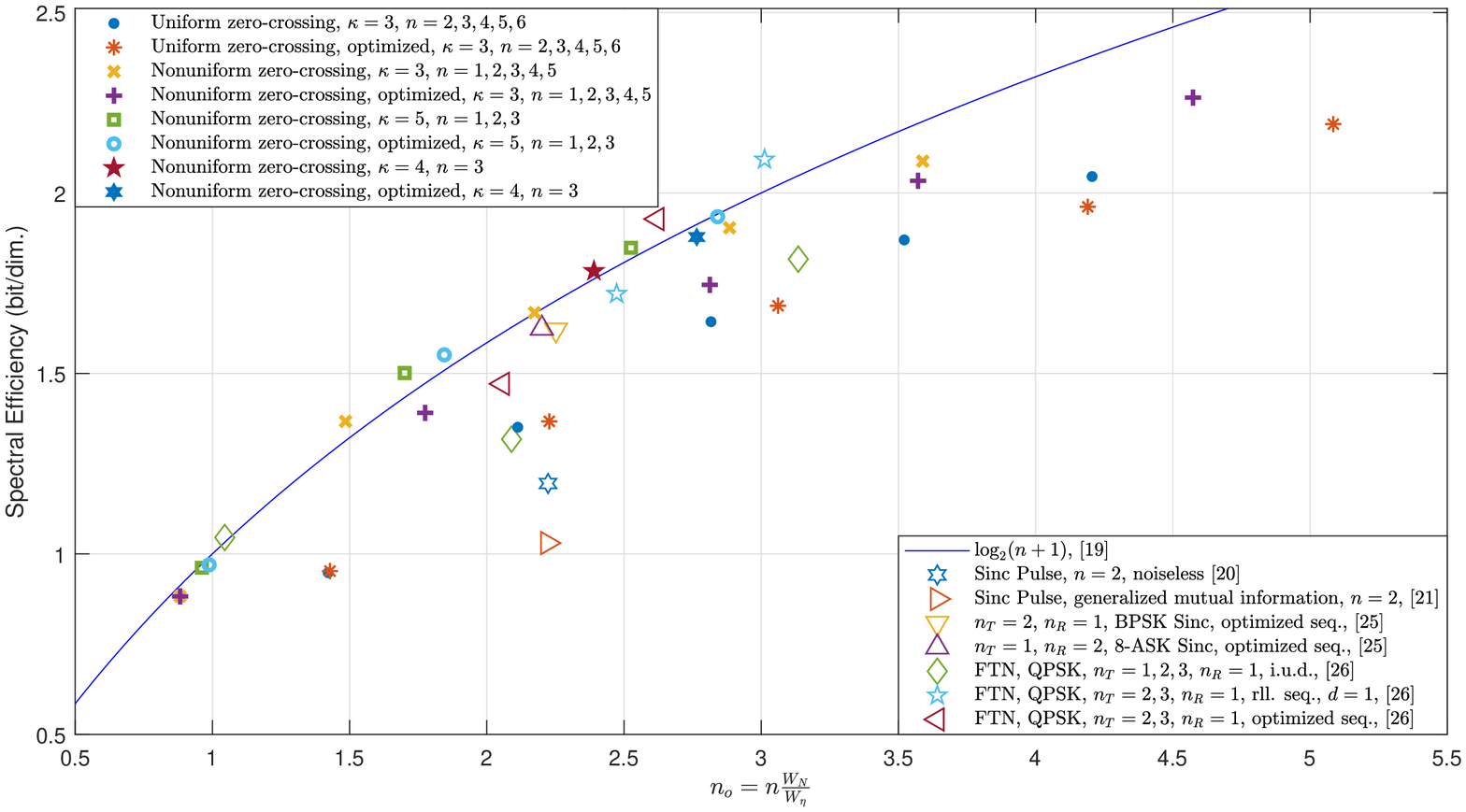}
}
\caption{{High-SNR limit of spectral efficiencies achieved by different schemes. Top: $\eta=0.95$; Bottom: $\eta=0.9$.  Effective oversampling factor $n_{\mathrm o}=n\frac{W_\mathrm N}{W_\eta}$. Oversampling factor for FTN (faster-than-Nyquist signaling) is $n=n_\mathrm T\cdot n_\mathrm R$ \cite{landauCL,landauJWCN}.}}
\label{fig_compare_capacity}
\end{figure*}

In Fig. \ref{fig_compare_capacity}, under different parameters, we show the high-SNR limits of spectral efficiencies achieved by our transceiver and those obtained in \cite{landauCL,landauJWCN} with respect to the fractional power containment bandwidth $W_\eta$, where we choose $\eta=0.95$ and $\eta=0.9$ as two examples.
For fair comparison, the results are shown with respect to the effective oversampling factor $n_{\mathrm o}=n\frac{W_\mathrm N}{W_\eta}$.
For clarity, only several representative results from those summarized in [\ref{landauJWCN}, Fig. 18, Fig. 19] are plotted, where we use $n_\mathrm{T}$ to denote the parameter for the faster-than-Nyquist (FTN) signaling \cite{RA09}
and $n_\mathrm{R}$ to denote the number of samples per symbol (see \cite{landauJWCN} for details).
Our transceiver is optimized using the heuristic procedure in Sec. \ref{SEFinite}, which increases the spectral efficiency but also enlarges $n_{\mathrm o}$ because $W_\eta$ is reduced.
We also plot $\log_2(n_{\mathrm o}+1)$ as a benchmark, which is the spectral efficiency achieved by the bandlimited process $S(t)$ under one-bit quantization and oversampling in the noiseless case \cite{Shamai1994}.
The benchmark reveals the performance limit of nonuniform zero-crossing pattern as $\kappa$ grows without bound and $\eta$ tends to one.
It is shown that the spectral efficiencies achieved by our transceiver increase roughly logarithmically with the sampling rate.
Specifically, by employing the nonuniform zero-crossing pattern, our transceiver achieves spectral efficiencies sufficiently close to the benchmark $\log_2(n_{\mathrm o}+1)$.
In general, our high-SNR performance results are comparable to those of \cite{landauCL,landauJWCN}.
We also provide finite-SNR performance comparisons under $\eta=0.9$ in Fig. \ref{compare_SNR_1} and Fig. \ref{compare_SNR_2}.
In each figure two schemes with similar high-SNR performance in Fig. \ref{fig_compare_capacity} are chosen for comparison.
In Fig. \ref{compare_SNR_1} the spectral efficiency of the scheme in \cite{landauJWCN} achieves the high-SNR limit earlier, while the schemes in Fig. \ref{compare_SNR_2} show similar performance.

\subsection{A BICM-ID Scheme and Performance}\label{sec_BICM}
\begin{figure}
\centering
\includegraphics[width=1\columnwidth]{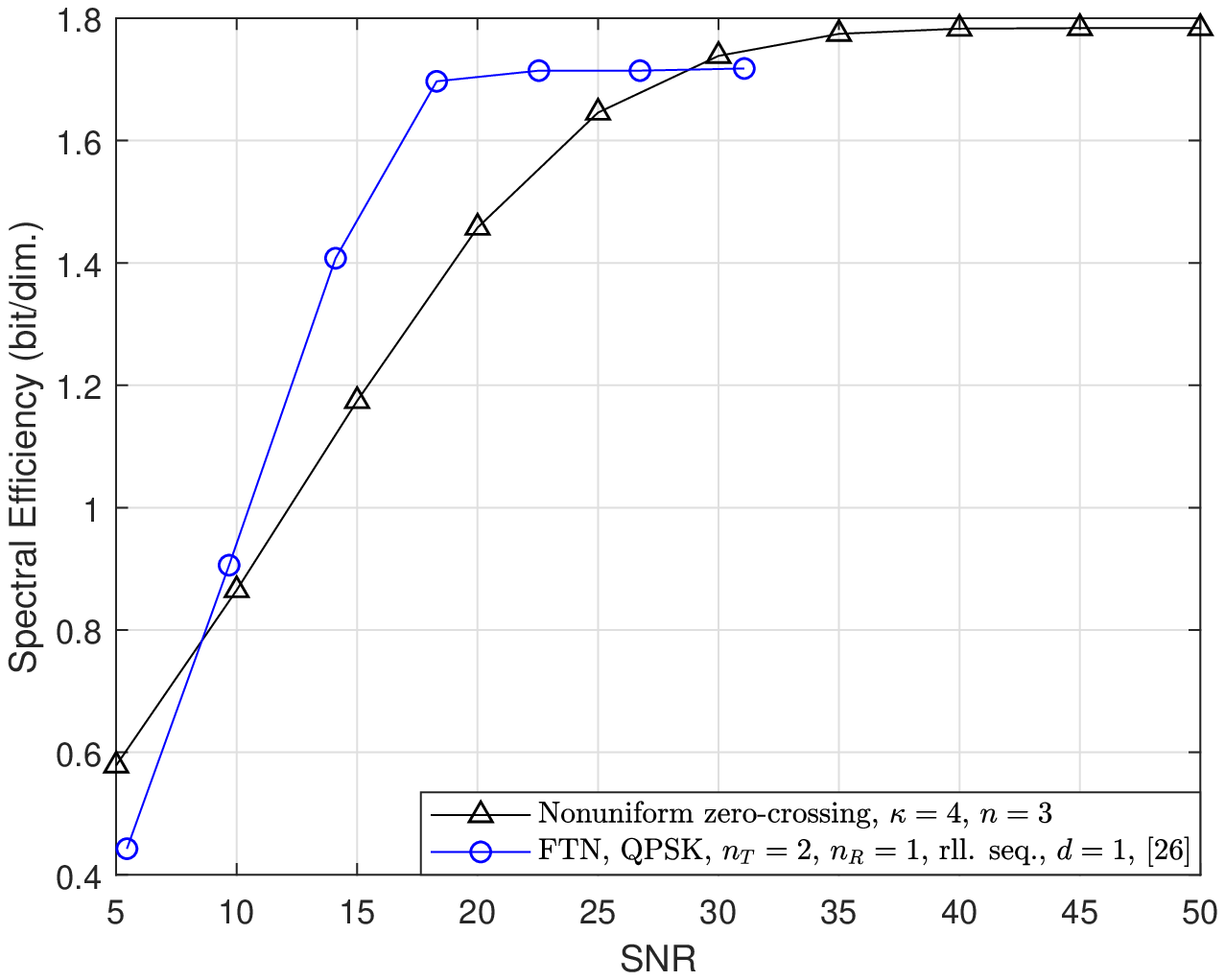}
\caption{{A finite-SNR comparison of two schemes in Fig. \ref{fig_compare_capacity}
under $\eta=0.9$ and $n_\mathrm o \approx 2.4$.}}
\label{compare_SNR_1}
\end{figure}
\begin{figure}
\centering
\includegraphics[width=1\columnwidth]{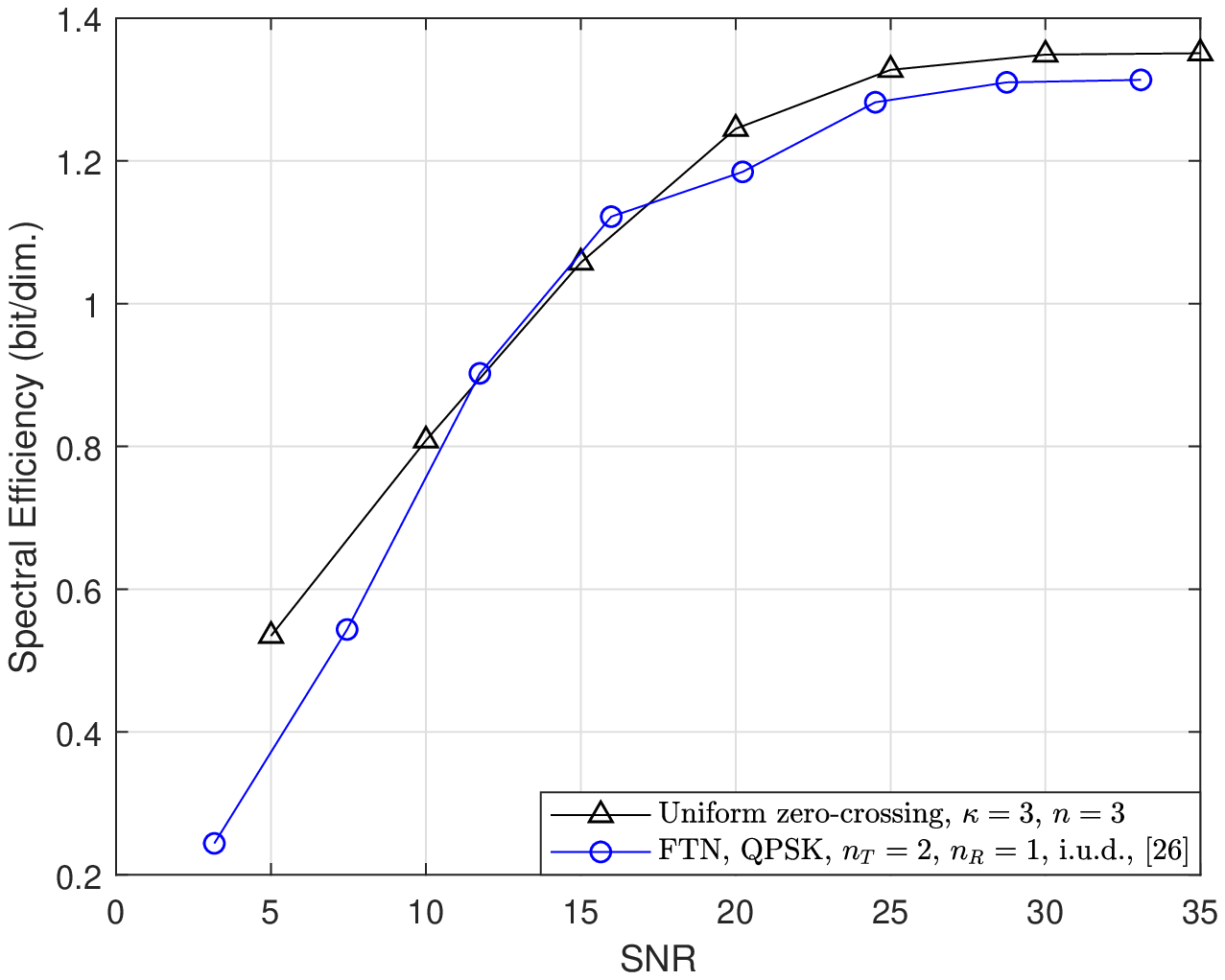}
\caption{{A finite-SNR comparison of two schemes in Fig. \ref{fig_compare_capacity}
under $\eta=0.9$ and $n_\mathrm o \approx 2.1$.}}
\label{compare_SNR_2}
\label{compare_SNR}
\end{figure}
Since our transceiver operates over a DMC, it is straightforward to incorporate standard coded modulation techniques.
As an example, we study error performance of a specific coded modulation scheme, namely, an LDPC coded bit-interleaved coded modulation with iterative decoding (BICM-ID) scheme, shown in Fig.~\ref{fig_system_con}, where the modulator and the demodulator include the building blocks of the transmitter and the receiver in Fig. \ref{fig:a}, respectively, except that the output of the demodulator is log-likelihood ratios (LLRs) rather than hard decisions.
Details of the scheme are given as follows.

\subsubsection{Transmitter}\label{subsec_encoder}
We employ a regular LDPC code where the parity check matrix is constructed following \cite{MacKay1999} and the generator matrix is correspondingly obtained by Gaussian elimination.
An essential problem is labeling, i.e., the mapping from coded bit tuples to symbols in $\mathcal U$.
According to the bijection $\mathcal U\to\mathcal G\to\mathcal S_{\mathcal G}$, if we let the size of $\mathcal G$ be an integer power of two (i.e., $\log_2 m=q\in \mathbb N$), we can define a labeling by a bijection from the set of all binary tuples of length $q$ to the set $\mathcal S_{\mathcal G}=\{\mathsf{sgn}(\mathbf g_u^\kappa), u=1,...,2^q\}$.
Fortunately, the performance loss due to such a restriction on $|\mathcal G|$ can be limited according to Sec. V-A.
However, even in this simple case, we cannot use the well-known Gray mapping which requires some regular distance property like QAM constellations.
Designing the optimal labeling is a difficult task.
In our example, we let $\kappa=3$, and assume that there are four zero-crossing positions in each Nyquist interval.
Thus we have $n^\kappa=64$ available waveform pairs when employing the uniform zero-crossing pattern (in this case $n=4$), or $(n+1)^\kappa-2^{\kappa-1}=60$ pairs when employing the nonuniform zero-crossing pattern (in this case $n=3$).
To reduce the occupied bandwidth, we exclude some of all the available waveform pairs following the method in Sec. V-A, and let $|\mathcal G|=64$ (i.e. 32 pairs).
We use a heuristic procedure to design the labeling scheme as follows, where $\mathbf a_u=[a_u[1],...,a_u[q]]$ is the bit tuple mapped to the sign sequence $\mathsf{sgn}(\mathbf g_u^\kappa)\in \mathcal S_{\mathcal G}$.

\begin{enumerate}[i)]
\item Since $\mathcal S_{\mathcal G}$ consists of $2^{q-1}$ pairs of antipodal sign sequences,
we can let each antipodal pair in all the bit tuples of length $q$ be mapped to an antipodal pair in $\mathcal S_{\mathcal G}$.
Moreover, we let the first $2^{q-1}$ bit tuples, $000000$ to $011111$, be mapped to the waveforms satisfying $g_u(0^+)>0$.
\item
For all $\mathsf{sgn}(\mathbf g_u^\kappa)$ included in $\mathcal S_{\mathcal G}$, the following Gray-like mapping is used:
When employing the uniform zero-crossing pattern, let
$00\Rightarrow [- + + +]$ and $01\Rightarrow [- - + +]$ for each Nyquist interval, and $11\Rightarrow [- - - +]$ and $10\Rightarrow [- - - -]$ in the 2nd and 3rd Nyquist intervals;
when employing the nonuniform zero-crossing pattern, let
$00\Rightarrow [+ + +]$ and $01\Rightarrow [- + +]$ for each Nyquist interval, and $11\Rightarrow [- - +]$ and $10\Rightarrow [- - -]$ in the 2nd and 3rd Nyquist intervals.
\item
Assume there are $m_1$ remaining bit tuples and the same number of remaining sign sequences after step ii).
Let $D(\mathbf a,\mathbf b)$ be the Hamming distance between two binary tuples $\mathbf a$ and $\mathbf b$.
Let $\{v_u\}$ be the set of the indices of the sequences $\mathbf g_v^\kappa$ such that $D\left(\mathsf{sgn}(\mathbf g_v^\kappa),\mathsf{sgn}(\mathbf g_u^\kappa)\right)=1$.
We then compute
\begin{equation}
D_\mathrm{sum}=\sum\limits_{u\in\mathcal U}\sum\limits_{v\in\{v_u\}} D(\mathbf a_v,\mathbf a_u)
\end{equation}
for all $m_1!$ possible mapping rules from the remaining bit tuples to the remaining sign sequences, and choose the one minimizing $D_\mathrm{sum}$ as our mapping rule.
\end{enumerate}

In step ii) above, note that Gray-like mappings are only used for a half of all available waveforms satisfying that the first zero-crossing occurs at the first or second possible positions.
This is because the waveforms of the other half are more likely to be excluded in the optimization of $|\mathcal G|$.
In fact, in our example employing the uniform zero-crossing pattern, 26 of 32 waveforms are excluded and we need to compute and compare only $6!=720$ different mapping schemes.

Moreover, introducing an interleaver can be helpful here, while in typical LDPC coded modulation systems it is not necessary.
We use an interleaver given by an $N\times q$ array, and we let $N=1024$ and $q=\log_2 m=6$ for example.
The coded bits are fed into the interleaver diagonally downwards from left to right, and read in a column-wise fashion; see Fig.~\ref{fig_interleaver}, where different grey levels indicate the positions corresponding to different LDPC codewords.
\begin{figure}
{\par\centering
\resizebox*{3in}{!}{\includegraphics{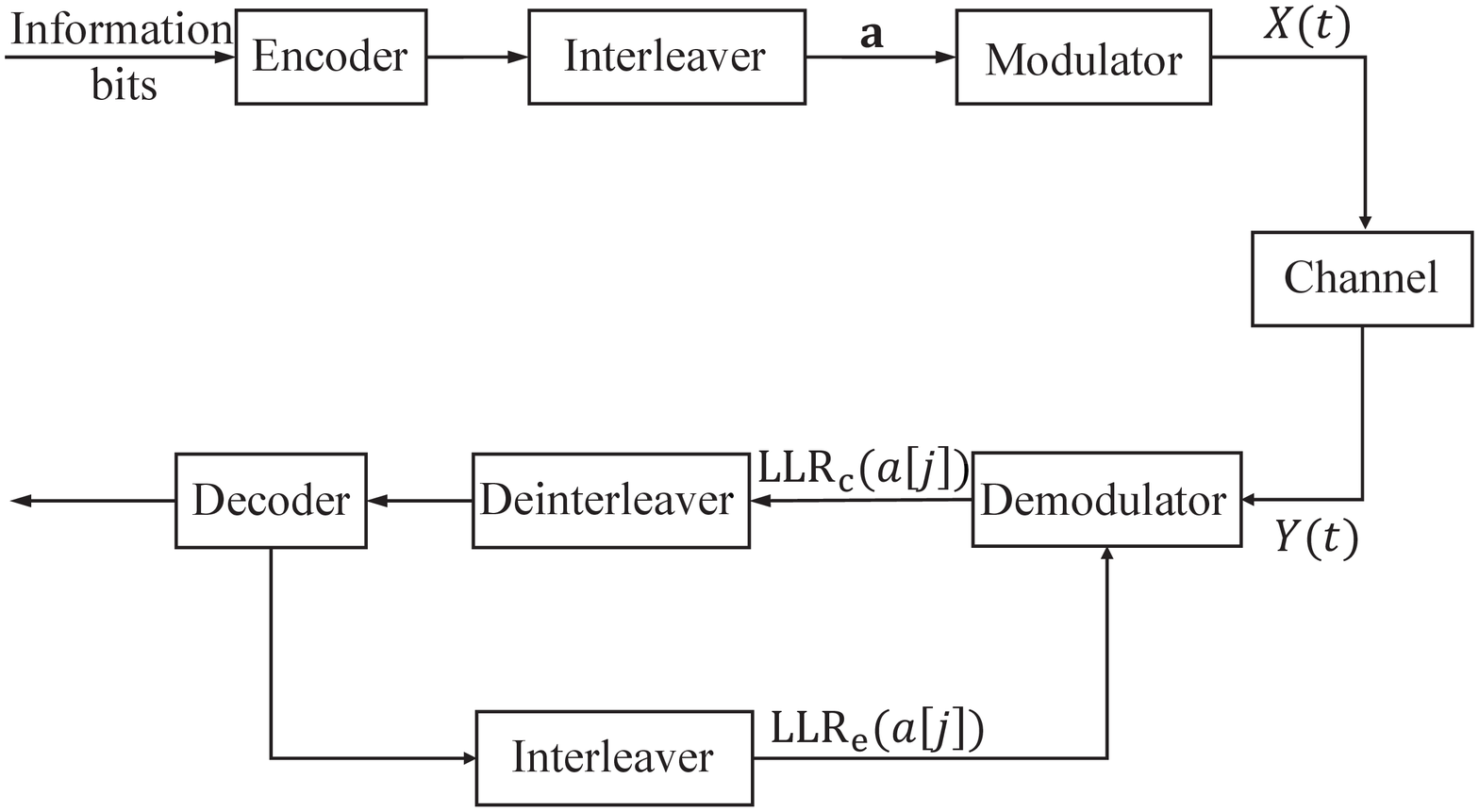}} \par}
\caption{System schematic illustration with BICM-ID.}
\label{fig_system_con}
\end{figure}
\begin{figure}
{\par\centering
\resizebox*{3in}{!}{\includegraphics{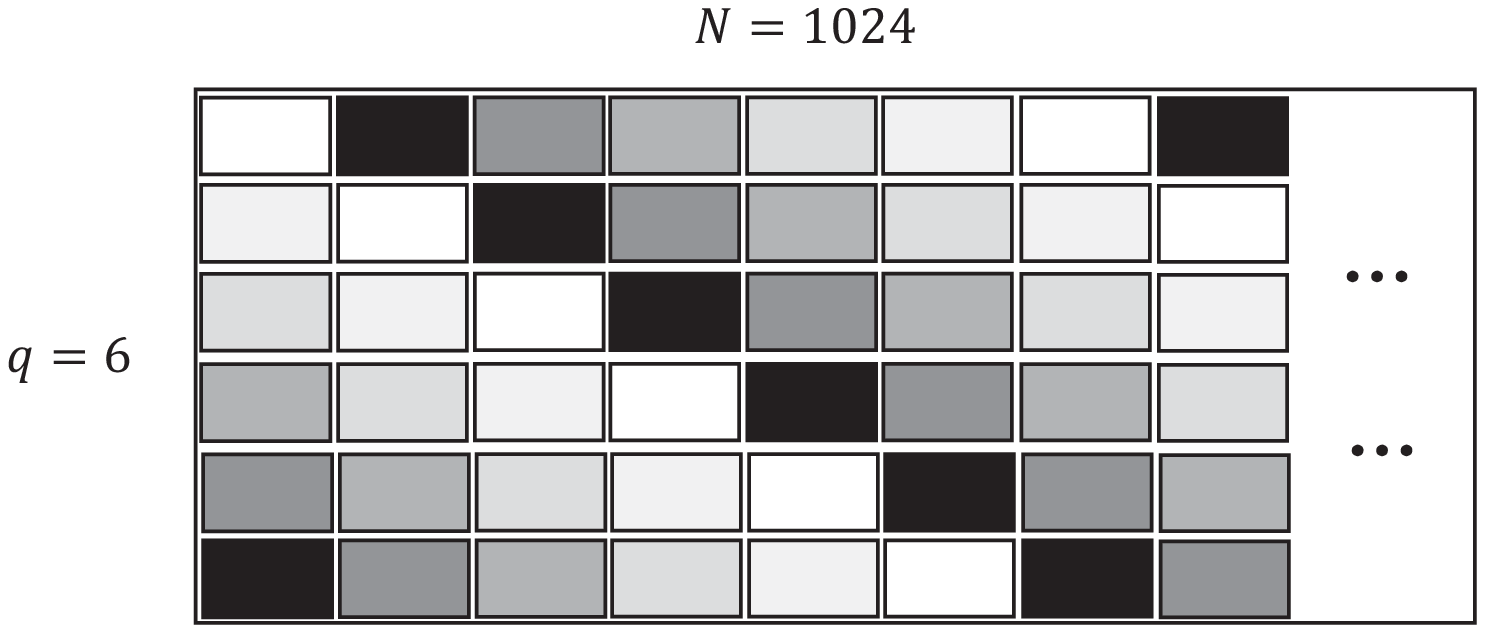}} \par}
\caption{The structure of the interleaver.}
\label{fig_interleaver}
\end{figure}

\subsubsection{Receiver}\label{subsec_decoder}

First, from the received signal $y(t)$, we obtain sign sequences $\{\mathbf b^\kappa\}$, each corresponding to a transmitted symbol. Then the receiver performs iterative decoding based on the sign sequences corresponding to a whole codeword as follows.
\begin{enumerate}[i)]
\item Initialization: For each coded bit tuple $\mathbf a=\left[a[1],...,a[q]\right]$, let the \emph{a prior} probability be $\Pr(a[j]=1)=\Pr(a[j]=0)=1/2$, $j=1,...,q$.
\item LLR calculation: based on the mapping rule $\{\mathbf a_u \rightarrow u\}$ and the observed sign sequence $\{\mathbf b^\kappa\}$, the demodulator calculates $\mathsf{LLR}_{\mathrm c}(a[j])=\log\frac{p_1}{p_0}$, where
\begin{equation} \label{eq_17}
p_a:=\sum\limits_{u\in \mathcal U_{j,a}}\left(p(\mathbf{b}^\kappa|u)
\prod\limits_{\substack{j^\prime\in\{1,...,q\},\\ j^\prime\neq j}}\Pr(a[j^\prime]=a_u[j^\prime])\right),
\end{equation}
and $\mathcal U_{j,a}$ is the set of $u$ such that the $j$-th bit of the bit tuple mapped to $g_u(t)$ satisfies $a_u[j]=a$.
\item Extrinsic information feedback and iteration:
According to $\mathsf{LLR}_{\mathrm c}(a[j])$, the LDPC decoder performs its own iterative decoding. After a certain number of iterations, the decoder (combined with the interleaver) generates the extrinsic information $\mathsf{LLR}_{\mathrm e}(a[j])=\log\frac{\Pr(a[j]=1)}{\Pr(a[j]=0)}$, which is fed back to the demodulator as the a prior information for the next iteration between the demodulator and the decoder.
Specifically, we have $\Pr(a_j=0)=\frac{\exp\left(\mathsf{LLR}_{\mathrm e}(a[j])\right)}{1+\exp\left(\mathsf{LLR}_{\mathrm e}(a[j])\right)}$ and $\Pr(a[j]=1)=\frac{1}{1+\exp\left(\mathsf{LLR}_{\mathrm e}(a[j])\right)}$.
\item Final decision: After a certain number of iterations, the final decision is made by the LDPC decoder.
\end{enumerate}

\subsubsection{Numerical Results}\label{subsec_mapping}
\begin{table*}[tbp]
\renewcommand{\cellset}
{\renewcommand{\arraystretch}{1}}
\centering
\caption{Parameters in BER Simulation}
\scalebox{1}{
\begin{tabular}{l|l}
\hline
Power Containment Factor&$\eta=0.95$\\\hline
Size of Waveform Set&$|\mathcal G|=m=64$\\\hline
Length of $g_u(t)$ (Nyquist Intervals)&$\kappa=3$\\\hline
Oversampling Factor&\makecell[cl]{Uniform zero-crossing pattern: $n=4$\\Nonuniform zero-crossing pattern: $n=3$} \\\hline
LDPC Code&Rate: 0.8125; Length: $1024$ \\\hline
Number of Iterations&Demodulator: 5; LDPC decoder: 50\\\hline
Spectral Efficiency Achieved&$1.1498$ bits/dim in Fig. \ref{fig_coded_BICM_ID_equal}, $1.2248$ bits/dim in Fig. \ref{fig_coded_BICM_ID_unequal}\\ \hline
\end{tabular}}
\end{table*}

We provide bit error rate (BER) performances of our transceiver employing LDPC coded modulation with simulation parameters in Table \uppercase\expandafter{\romannumeral3}.
For comparison, the results are provided under different settings:
i) employing the uniform zero-crossing pattern (Fig. \ref{fig_coded_BICM_ID_equal}) or the nonuniform one (Fig. \ref{fig_coded_BICM_ID_unequal});
ii) employing the proposed mapping rule or a random mapping rule (i.e., a randomly generated bijection from the set of all binary tuples of length $q$ to $\mathcal U$);
iii) without (Fig. \ref{BER_equal_uninter} and Fig. \ref{BER_unequal_uninter}) or with (Fig. \ref{BER_equal_inter} and Fig. \ref{BER_unequal_inter}) an interleaver.
In general, the results demonstrate that our transceiver can achieve spectral efficiencies larger than one bit per dimension at moderate SNRs with sufficiently low BERs.
It is shown that the performance under the nonuniform zero-crossing pattern and an oversampling factor $n=3$ is similar to that under uniform zero-crossing pattern and $n=4$.
Compared to the case of random mapping, performances can be notably improved using the mapping schemes found by our heuristic procedure.
Moreover, introducing an interleaver improves the performance, and the gain is larger when the random mapping is used.
We note that the interleaver enlarges the decoding delay, but it does not increase the decoding complexity.

\begin{figure*}
\centering
\subfigure[Performance without interleaver.]
{
\begin{minipage}[t]{0.48\linewidth} \centering \label{BER_equal_uninter}
\includegraphics[width=1\columnwidth]{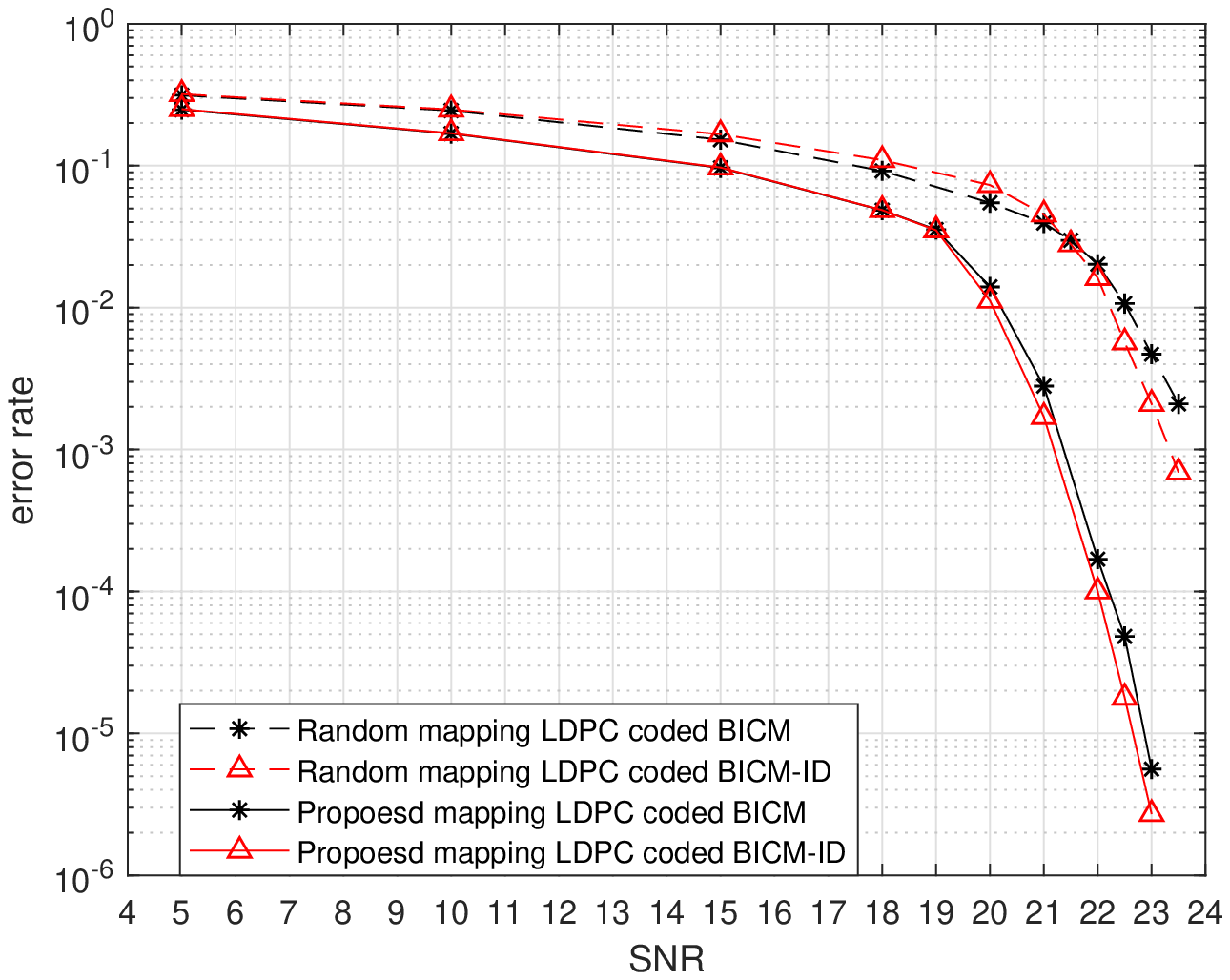}
\end{minipage}
}
\subfigure[Performance with proposed interleaver.]
{
\begin{minipage}[t]{0.48\linewidth} \centering \label{BER_equal_inter}
\includegraphics[width=1\columnwidth]{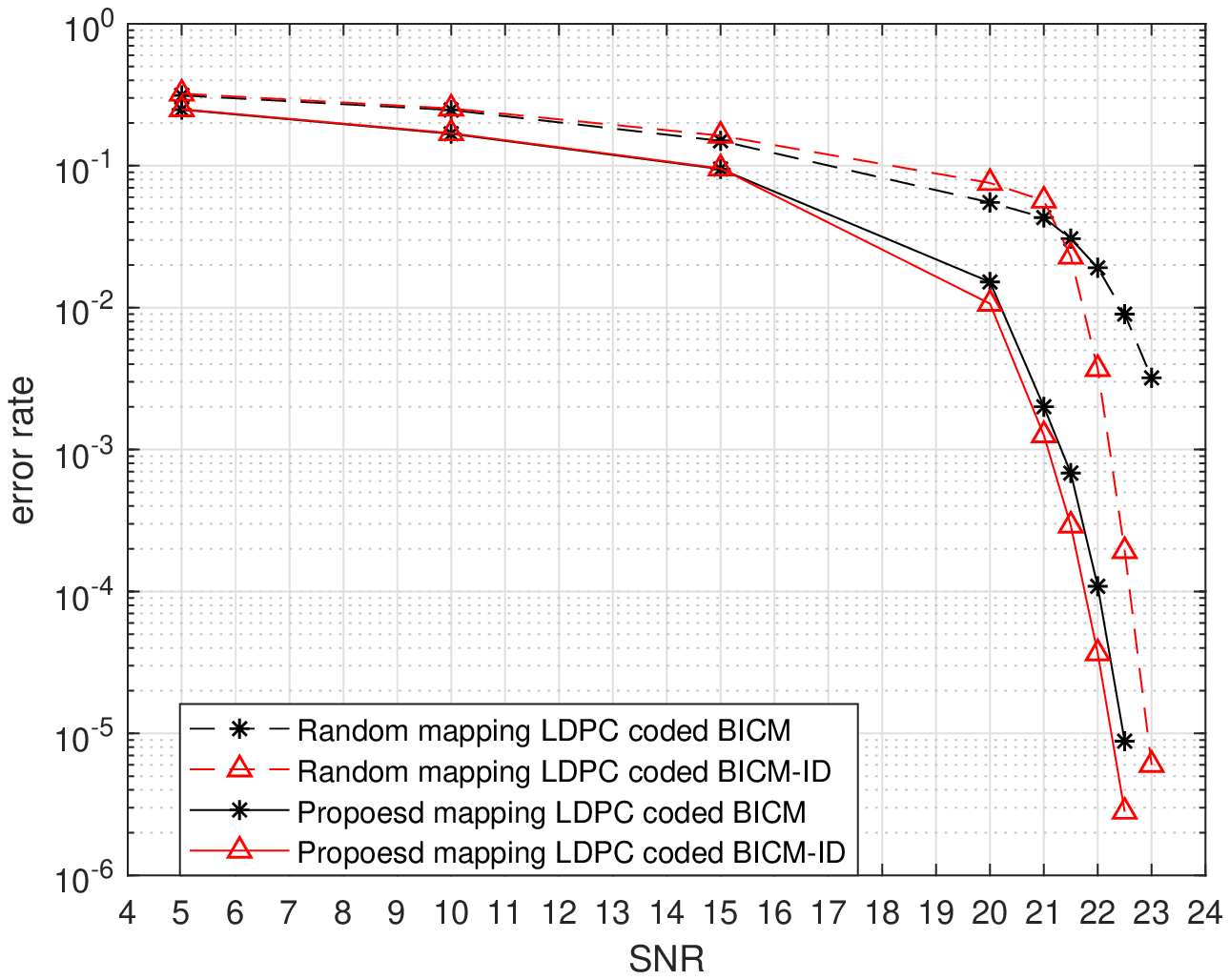}
\end{minipage}
}
\caption{BER performance of our transceiver with uniform zero-crossing pattern and LDPC coded modulation.}
\label{fig_coded_BICM_ID_equal}
\end{figure*}
\begin{figure*}
\centering
\subfigure[Performance without interleaver.]
{
\begin{minipage}[t]{0.48\linewidth} \centering \label{BER_unequal_uninter}
\includegraphics[width=1\columnwidth]{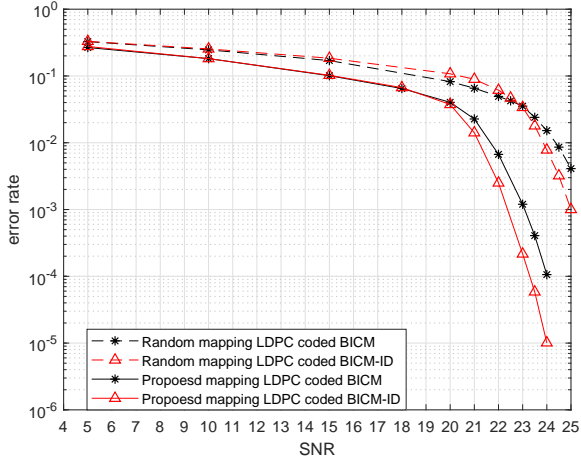}
\end{minipage}
}
\subfigure[Performance with proposed interleaver.]
{
\begin{minipage}[t]{0.48\linewidth} \centering \label{BER_unequal_inter}
\includegraphics[width=1\columnwidth]{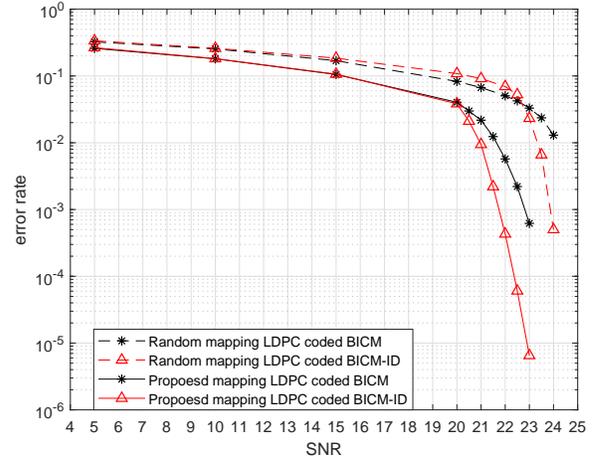}
\end{minipage}
}
\caption{BER performance of our transceiver with nonuniform zero-crossing pattern and LDPC coded modulation.}
\label{fig_coded_BICM_ID_unequal}
\end{figure*}

\section{Conclusion}\label{conclusion}
In this work, we propose a transceiver structure for bandlimited communication under the condition that a one-bit quantizer is employed at the receiver.
The transmitter employs a finite set of time-limited and approximately bandlimited waveforms to construct the channel input, and the receiver employs an integrate-and-dump receiver which observes the channel output faster than the Nyquist sampling rate.
The transceiver structure reduces the detection complexity and enables application of standard capacity-achieving coded modulation techniques.
Both transceiver information theoretic analysis and error performance simulations demonstrate that our transceiver can achieve spectral efficiencies larger than one bit per dimension with acceptable receiver complexity at moderate to high SNR, thereby confirming that considerable performance gains can be obtained by oversampling.

\appendices
\section{Proof of Proposition 1}

We prove the case $i=1$, then the general case can be proved by recursion.
When $i=1$, we have \cref{st,st1,sit},
\begin{figure*}
\begin{equation}
\label{st}
s(\boldsymbol{\delta},t)=(t-\delta_0)\lim_{K\to\infty}\prod\limits_{k=1}^{K}\left(1-\frac{t}{k+\delta_k}\right)\left(1-\frac{t}{-k+\delta_{-k}}\right).
\end{equation}
\end{figure*}
\begin{figure*}
\begin{align}
\label{st1}
s\left(\boldsymbol{\delta}_{(1)},t\right)=(t-\delta_1)\lim_{K\to\infty}\prod\limits_{k=1}^{K}\left(1-\frac{t}{k+\delta_{k+1}}\right)\left(1-\frac{t}{-k+\delta_{-k+1}}\right).
\end{align}
\end{figure*}
\begin{figure*}
\begin{flalign}
\label{sit}
s\left(\boldsymbol{\delta}_{(1)},t-1\right)&=(t-1-\delta_1)\lim_{K\to\infty}\prod\limits_{k=1}^{K}\left(1-\frac{t-1}{k+\delta_{k+1}}\right)\left(1-\frac{t-1}{-k+\delta_{-k+1}}\right)\notag\\
&=\frac{t-\delta_0}{1-\delta_0}(t-1-\delta_1)\lim_{K\to\infty}\Bigg(\prod\limits_{k=1}^{K}\frac{k+1+\delta_{k+1}}{k+\delta_{k+1}}\left(1-\frac{t}{k+1+\delta_{k+1}}\right)\notag\\
&\mspace{225mu}\prod\limits_{k=2}^{K}\frac{-k+1+\delta_{-k+1}}{-k+\delta_{-k+1}}\left(1-\frac{t}{-k+1+\delta_{-k+1}}\right)\Bigg)\notag\\
&=-\frac{1+\delta_1}{1-\delta_0}(t-\delta_0)\lim_{K\to\infty}\Bigg(\prod\limits_{k=1}^{K}\left(1+\frac{1}{k+\delta_{k+1}}\right)\prod\limits_{k=2}^{K}\left(1-\frac{1}{k-\delta_{-k+1}}\right)\notag\\
&\mspace{205mu} \prod\limits_{k=1}^{K+1}\left(1-\frac{t}{k+\delta_{k}}\right)\prod\limits_{k=1}^{K-1}\left(1-\frac{t}{-k+\delta_{-k}}\right)\Bigg)\notag\\
&=-c_1\left(t-\delta_0\right)\lim_{K\to\infty}\left(\prod\limits_{k=1}^{K+1}\left(1-\frac{t}{k+\delta_{k}}\right)\prod\limits_{k=1}^{K-1}\left(1-\frac{t}{-k+\delta_{-k}}\right)\right).
\end{flalign}
\end{figure*}
where $c_1$ is a constant satisfying
\begin{align}
c_1&=\frac{1+\delta_1}{1-\delta_0}\lim_{K\to\infty}\bigg(\prod\limits_{k=1}^{K}\left(1+\frac{1}{k+\delta_{k+1}}\right)\notag\\
&\mspace{135mu}\prod\limits_{k=2}^{K}\left(1-\frac{1}{k-\delta_{-k+1}}\right)\bigg)\notag\\
&<\frac{1+\frac{1}{2}}{1-\frac{1}{2}}\left(1+\frac{1}{1-\frac{1}{2}}\right)\lim_{K\to\infty}\prod\limits_{k=2}^{K}\left(1+\frac{1}{k-\frac{1}{2}}\right)\notag\\
&\mspace{255mu}\left(1-\frac{1}{k+\frac{1}{2}}\right)\notag\\
&=9 \notag,
\end{align}
\begin{align}
c_1&>\frac{1-\frac{1}{2}}{1+\frac{1}{2}}\lim_{K\to\infty}\left(1+\frac{1}{1+\frac{1}{2}}\right)\prod\limits_{k=2}^{K}\left(1+\frac{1}{k+\frac{1}{2}}\right)\notag\\
&\mspace{255mu}\left(1-\frac{1}{k-\frac{1}{2}}\right)\notag\\
&=\frac{5}{9}\lim_{K\to\infty}\prod\limits_{k=2}^{K}\left(1-\frac{2}{k^2-\frac{1}{4}}\right)\notag=\rm constant>0,
\end{align}
because $\sum_{k=1}^{\infty}\frac{2}{k^2-\frac{1}{4}}$ converges.
The proof is completed by comparing (\ref{st}) and (\ref{sit}).

\section{Proof of Proposition 2}
The following proof is an extension of a proof in [\ref{Lapidoth2017}, pp.277-278], in which the average autocovariance function of a pulse amplitude modulation signal is derived.
For every $t,\tau\in\mathbb R$, we have (\ref{EX}),
\begin{figure*}
\begin{align}
\label{EX}
\mathrm{E}\left[X(t)X(t+\tau)\right]&= \mathrm{E}\left[\left(\sum_{i}G_{U_i}\left(t-i\kappa T_{\mathrm N}\right)\right)\left(\sum_{i'}G_{U_{i'}}\left(t+\tau-i'\kappa T_{\mathrm N}\right)\right)\right]\notag\\
&=\sum_i\sum_{i'}\mathrm{E}\left[\left(\sum_{u=1}^{m}\mathds{1}_{U_i}(u)g_u\left(t-i\kappa T_{\mathrm N}\right)\right)\left(\sum_{u'=1}^{m}\mathds{1}_{U_{i'}}(u')g_{u'}\left(t+\tau-i'\kappa T_{\mathrm N}\right)\right)\right]\notag\\
&=\sum_i\sum_{i'}\sum_{u=1}^{m}\sum_{u'=1}^{m}\mathrm{E}\left[\mathds{1}_{U_i}(u)\mathds{1}_{U_{i'}}(u')\right]g_u\left(t-i\kappa T_{\mathrm N}\right)g_{u'}\left(t+\tau-i'\kappa T_{\mathrm N}\right)\notag\\
&=\underbrace{\frac{1}{m^2}\sum_{i,i',i\neq i'}\sum_{u=1}^{m}\sum_{u'=1}^{m}g_u\left(t-i\kappa T_{\mathrm N}\right)g_{u'}\left(t+\tau-i'\kappa T_{\mathrm N}\right)}\limits_{=0}\notag\\
&\mspace{20mu}+\frac{1}{m}\sum_{i}\sum_{u=1}^{m}g_u\left(t-i\kappa T_{\mathrm N}\right)g_{u}\left(t+\tau-i\kappa T_{\mathrm N}\right)\notag\\
&=\frac{1}{m}\sum_{i}\sum_{u=1}^{m}g_u\left(t-i\kappa T_{\mathrm N}\right)g_{u}\left(t+\tau-i\kappa T_{\mathrm N}\right),
\end{align}
\end{figure*}
where the last equality follows from the fact that $\mathcal G$ consists of antipodal pairs.
To derive the average autocovariance function of $X(t)$, we consider the integral over an arbitrary interval of length $\kappa T_{\mathrm N}$ as
\begin{align}
\label{intT}
&\int_{\delta}^{\delta+\kappa T_{\mathrm N}}\mathrm{E}\left[X(t)X(t+\tau)\right]dt\notag\\
=&\frac{1}{m}\int_{\delta}^{\delta+\kappa T_{\mathrm N}}\sum_{i}\sum_{u=1}^{m}g_u\left(t-i\kappa T_{\mathrm N}\right)g_{u}\left(t+\tau-i\kappa T_{\mathrm N}\right)dt\notag\\
=&\frac{1}{m}\sum_{i}\sum_{u=1}^{m}\int_{\delta}^{\delta+\kappa T_{\mathrm N}}g_u\left(t-i\kappa T_{\mathrm N}\right)g_{u}\left(t+\tau-i\kappa T_{\mathrm N}\right)dt\notag\\
=&\frac{1}{m}\sum_{u=1}^{m}\sum_{i}\int_{\delta-i\kappa T_{\mathrm N}}^{\delta-(i-1)\kappa T_{\mathrm N}}g_u\left(t\right)g_{u}\left(t+\tau\right)dt\notag\\
=&\frac{1}{m}\sum_{u=1}^{m}\int_{-\infty}^{\infty}g_u\left(t\right)g_{u}\left(t+\tau\right)dt\notag\\
=&\frac{1}{m}\sum_{u=1}^{m}{\mathsf A}_u(\tau),
\end{align}
where $\mathsf{A}_u(\tau)=\int_{-\infty}^{\infty}g_u(t+\tau)g_u(t)dt,\mspace{4mu} \tau\in\mathbb R$ is the autocorrelation of $g_u(t)$,
which satisfies $\hat{\mathsf{A}}_u=\left|\hat{g}_u\right|^2$.
Since (\ref{intT}) holds for every interval of length $\kappa T_{\mathrm N}$, we obtain
\begin{align}
\mathsf{A}_{\mspace{-2mu}X\mspace{-4mu}X}(\tau)&=\lim\limits_{T\to\infty}\frac{1}{2T}\int_{-T}^{T}\mathrm{cov}\left[X(t),X(t+\tau)\right]dt\notag\\
&=\frac{1}{m\kappa T_{\mathrm N}}\sum_{u=1}^{m}{\mathsf A}_u(\tau),
\end{align}
where the first equality follows from the fact that $X(t)$ has zero mean.
Taking the Fourier transform of $\mathsf{A}_{\mspace{-2mu}X\mspace{-4mu}X}(\tau)$ completes the proof of (\ref{eq_pf}).
Finally, (\ref{PE}) can be obtained by combining (\ref{eq_pf}) and the fact $\int_{-\infty}^{\infty}g_u^2(t)dt=\int_{-\infty}^{\infty}|\hat{g}_u(f)|^2df$.

\ifCLASSOPTIONcaptionsoff
  \newpage
\fi


\begin{thebibliography}{1}
\bibitem{SPM-bottleneck09}
J. Singh, S. Ponnuru, and U. Madhow, ``Multi-Gigabit communication: the ADC bottleneck,'' in \emph{Proc. IEEE Int. Conf. Ultra-Wideband (ICUWB)}, Vancouver, BC, Canada, Sep. 2009, pp. 22--27.

\bibitem{Fettweis19}
G. P. Fettweis, M. D{\"o}rpinghaus, J. Castrillon, A. Kumar, C. Baier, K. Bock, F. Ellinger, A. Fery, F. H. P. Fitzek, H. H{\"a}rtig, K. Jamshidi, T. Kissinger, W. Lehner, M. Mertig, W. E. Nagel, G. T. Nguyen, D. Plettemeier, M. Schr{\"o}ter, and T. Strufe, ``Architecture and advanced electronics pathways toward highly adaptive energy-efficient computing,'' \emph{Proc. IEEE}, vol. 107, no. 1, pp. 204--231, Jan. 2019.

\bibitem{Walden1999}
R. H. Walden, ``Analog-to-digital converter survey and analysis,'' \emph{IEEE J. Select. Areas Commun.}, vol. 17, no. 4, pp. 539--550, Apr. 1999.

\bibitem{Murmann15}
B. Murmann, ``The race for the extra decibel: A brief review of current ADC performance trajectories,'' \emph{IEEE Solid-State Circuits Mag.}, vol. 7, no. 3, pp. 58--66, Summer 2015.

\bibitem{Nossek}
J. A. Nossek and M. T. Ivrla{\v c}, ``Capacity and coding for quantized MIMO systems,'' in \emph{Proc. Intl. Conf. Wireless Commun. Mobile Computing (IWCMC)}, 2006, Vancouver, Canada, pp. 1387--1391.

\bibitem{singh2009limits}
J. Singh, O. Dabeer, and U. Madhow, ``On the limits of communication with low-precision analog-to-digital conversion at the receiver,'' \emph{IEEE Trans. Commun.}, vol. 57, no. 12, pp. 3629--3639, Dec. 2009.

\bibitem{zeitler2012low-precision}
G. Zeitler, A. C. Singer, and G. Kramer, ``Low-precision A/D conversion for maximum information rate in channels with memory,'' \emph{IEEE Trans. Commun.}, vol. 60, no. 9, pp. 2511--2521, Sep. 2012.

\bibitem{mo2015capacity}
J. Mo and R. W. Heath, ``Capacity analysis of one-bit quantized MIMO systems with transmitter channel state information,'' \emph{IEEE Trans. Signal Process.}, vol. 63, no. 20, pp. 5498--5512, Oct. 2015.

\bibitem{Zhang16}
N. Liang and W. Zhang, ``Mixed-ADC massive MIMO,'' \emph{IEEE J. Sel. Areas Commun.}, vol. 34, no. 4, pp. 983--997, Apr. 2016.

\bibitem{CMH16}
J. Choi, J. Mo, and R. W. Heath, ``Near maximum-likelihood detector and channel estimator for uplink multiuser massive MIMO systems with one-bit ADCs,'' \emph{IEEE Trans. Commun.}, vol. 64, no. 5, pp. 2005--2018, May 2016.

\bibitem{Studer16}
C. Studer and G. Durisi, ``Quantized massive MU-MIMO-OFDM uplink,'' \emph{IEEE Trans. Commun.}, vol. 64, no. 6, pp. 2387--2399, Jun. 2016.

\bibitem{Rini17ITW}
S. Rini, L. Barlett, E. Erkip, and Y. C. Eldar, ``A general framework for MIMO receivers with low-resolution quantization,'' in \emph{Proc. IEEE Inf. Theory Workshop (ITW)}, Kaohsiung, Taiwan, Nov. 2017, pp. 599--603.

\bibitem{Abbas17}
W. B. Abbas, F. Gomez-Cuba, and M. Zorzi, ``Millimeter wave receiver efficiency: A comprehensive comparison of beamforming schemes with low resolution ADCs,'' \emph{IEEE Trans. Wirel. Commun.}, vol. 16, no. 12, pp. 8131--8146, Dec. 2017.

\bibitem{Rini18ISIT}
A. Khalili, S. Rini, L. Barletta, E. Erkip, and Y. C. Eldar, ``On MIMO channel capacity with output quantization constraints,'' in \emph{Proc. 2018 IEEE Int. Symp. Inf. Theory (ISIT)}, Jun. 2018, pp. 1355--1359.

\bibitem{VO79}
A. J. Viterbi and J. K. Omura, \emph{Principles of Digital Communication and Coding}, McGraw-Hill, 1979.

\bibitem{KochLapidoth}
T. Koch and A. Lapidoth, ``At low SNR, asymmetric quantizers are better,'' \emph{IEEE Trans. Inf. Theory}, vol. 59, no. 9, pp. 5421--5445, Sep. 2013.

\bibitem{FU98}
G. D. Forney and G. Ungerboeck, ``Modulation and coding for linear Gaussian channels,'' \emph{IEEE Trans. Inf. Theory}, vol. 44, no.6, pp. 2384--2415, Oct. 1998.

\bibitem{Gilbert1993}
E. N. Gilbert, ``Increased information rate by oversampling,'' \emph{IEEE Trans. Inf. Theory}, vol. 39, no. 6, pp. 1973--1976, Nov. 1993.

\bibitem{Shamai1994}
S. Shamai (Shitz), ``Information rates by oversampling the sign of a bandlimited process,'' \emph{IEEE Trans. Inf. Theory}, vol. 40, no. 4, pp. 1230--1236, July 1994.

\bibitem{KochLapidoth1}
T. Koch and A. Lapidoth, ``Increased capacity per unit-cost by oversampling,'' in \emph{Proc. IEEE 26th Conv. Elect. Electron. Eng. Israel}, Eilat, Israel, Nov. 2010, pp. 684--688.

\bibitem{Zhang2012}
W. Zhang, ``A general framework for transmission with transceiver distortion and some applications,'' \emph{IEEE Trans. Commun.}, vol. 60, no. 2, pp. 384--399, Feb. 2012.

\bibitem{Krone2012}
S. Krone and G. Fettweis, ``Communications with 1-bit quantization and oversampling at the receiver: Benefiting from inter-symbol-interference,'' in \emph{Proc. 2012 IEEE 23rd Int. Symp. Pers. Indoor Mobile Radio Commun.}, 2012, pp. 2408--2413.

\bibitem{Halsig2014}
T. H{\"a}lsig, L. Landau, and G. Fettweis, ``Spectral efficient communications employing 1-bit quantization and oversampling at the receiver,'' in \emph{Proc. IEEE 80th Veh. Technol. Conf. (VTC)}, Vancouver, BC, Canada, 2014.

\bibitem{Singh2015}
G. Singh, L. Landau, and G. Fettweis, ``Finite length reconstructible ASK-sequences received with 1-bit quantization and oversampling,'' in \emph{Proc. 10th Int. ITG Conf. Syst. Commun. Coding (SCC)}, Hamburg, Germany, 2015.

\bibitem{landauCL}
L. Landau, M. D{\"o}rpinghaus, and G. P. Fettweis, ``1-bit quantization and oversampling at the receiver: Communication over bandlimited channels with noise,'' \emph{IEEE Commun. Lett.}, vol. 21, no. 5, pp. 1007--1010, May 2017.

\bibitem{landauJWCN}
\label{landauJWCN}
L. T. N. Landau, M. D{\"o}rpinghaus, and G. P. Fettweis, ``1-bit quantization and oversampling at the receiver: Sequence-based communication,'' \emph{EURASIP J. Wirel. Commun. Netw.}, vol. 2018, no. 1, pp. 83, Apr. 2018.

\bibitem{Bender}
S. Bender, M. D{\"o}rpinghaus, and G. Fettweis, ``On the achievable rate of bandlimited continuous-time 1-bit quantized AWGN channels,'' in \emph{Proc. 2017 IEEE Int. Symp. Inf. Theory (ISIT)}, Aachen, Germany, 2015, pp. 2083--2087.

\bibitem{landauTWC}
L. T. N. Landau, M. D{\"o}rpinghaus, R. C. de Lamare, and G. P. Fettweis, ``Achievable rate with 1-bit quantization and oversampling using continuous phase modulation based sequences,'' \emph{IEEE Trans. Wirel. Commun.}, vol. 17, no. 10, pp. 7080--7095, Oct. 2018.

\bibitem{GBLV17}
A. Gokceoglu, E. Bj{\"o}rnson, E. G. Larsson, and M. Valkama, ``Spatio-temporal waveform design for multiuser massive MIMO downlink with 1-bit receivers,'' \emph{IEEE J. Sel. Topics Signal Process.}, vol. 11, no. 2, pp. 347--362, Mar. 2017.

\bibitem{UY18}
A. B. {\"U}{\c c}{\"u}nc{\"u} and A. {\"O}. Y{\i }lmaz, ``Oversampling in one-bit quantized massive MIMO systems and performance analysis,'' \emph{IEEE Trans. Wirel. Commun.}, vol. 17. no. 12, pp. 7952--7964, Dec. 2018.

\bibitem{Fettweis}
G. Fettweis, M. D{\"o}rpinghaus, S. Bender, L. Landau, P. Neuhaus, and M. Schl{\"u}ter, ``Zero crossing modulation for communcation with temporally oversampled 1-bit quantization,'' in \emph{Proc. 53rd Asilomar Conf. Signals, Syst. Comput.}, Asilomar, CA, USA, Nov. 2019, pp. 207--214.

\bibitem{Alencar19}
R. R. M. de Alencar, L. T. N. Landau, and R. C. de Lamare, ``Continuous phase modulation with 1-bit quantization and oversampling using iterative detection and decoding,'' in \emph{Proc. 53rd Asilomar Conf. on Signals, Syst. Comput.}, Asilomar, CA, USA, Nov. 2019, pp. 1729--1733.

\bibitem{DC}
A. B. Carlson and P. B. Crilly, \emph{Communication Systems}, 5th ed., Mc Graw Hill, 2009.

\bibitem{P}
S. J. Henely and W. D. Walby, ``Differential phase-shift keying demodulator,'' U.S. patent 4,896,336, Jan. 1990, Rockwell Internatioal Corporation, El Segundo, CA, USA.

\bibitem{Eldar13}
Y. Chen, Y. C. Eldar, and A. J. Goldsmith, ``Shannon meets Nyquist: Capacity of sampled Gaussian channels,'' \emph{IEEE Trans. Inf. Theory}, vol. 59, no. 8, pp. 4889--4914, Aug. 2013.

\bibitem{WTT77}
G. L. Wise, A. P. Traganitis, and J. B. Thomas, ``The effect of a memoryless nonlinearity on the spectrum of a random process,'' \emph{IEEE Trans. Inf. Theory}, vol. 23, no, 1, pp. 84--89, Jan. 1977.

\bibitem{Eldar}
Y. C. Eldar, \emph{Sampling Theory: Beyond Bandlimited Systems}, Cambridge University Press, 2015.

\bibitem{Wong}
\label{Wong}
E. Wong, ``Recent progress in stochastic processes--A survey'', \emph{IEEE Trans. Inf. Theory}, vol. 19, no. 3, pp. 262--275, May 1973.

\bibitem{Requicha80}
A. A. G. Requicha, ``The zeros of entire functions: Theory and engineering applications,'' \emph{Proc. IEEE}, vol. 68, no. 3, pp. 308--328, Mar. 1980.

\bibitem{MLR2}
D. M. V. Melo, L. T. N. Landau, and R. C. de Lamare, ``Zero-crossing precoding with MMSE criterion for channels with 1-bit quantization and oversampling," in \emph{Proc. 24th Int. ITG Wksp. Smart Antennas}, Hamburg, Germany, Feb 2020.

\bibitem{MLR1}
D. M. V. Melo, L. T. N. Landau, and R. C. de Lamare, ``Zero-crossing precoding with maximum distance to the decision threshold for channels with 1-bit quantization and oversampling," in \emph{Proc. IEEE Int. Conf. Acoust. Speech \& Signal Process. (ICASSP)}, Barcelona, Spain, May 2020, pp. 5120-5124.

\bibitem{GK17}
H. Ghozlan, G. Kramer, ``Models and information rates for Wiener phase noise channels,'' \emph{IEEE Trans. Inf. Theory}, vol. 63, no. 4, pp. 2376--2393, Apr. 2017.

\bibitem{Jacobsson17DAC}
S. Jacobsson, G. Durisi, M. Coldrey, T. Goldstein, and C. Studer, ``Quantized precoding for massive MU-MIMO,'' \emph{IEEE Trans. Commun.}, vol. 65, no. 11, pp. 4670--4684, Nov. 2017.

\bibitem{Dutta20}
S. Dutta, A. Khalili, E. Erkip, and S. Rangan, ``Capacity bounds for communication systems with quantization and spectral constraints,'' in \emph{Proc. 2020 IEEE Int. Symp. Inf. Theory (ISIT)}, Virtual Conference, June 2020, pp. 2056--2061.

\bibitem{Lapidoth2017}
\label{Lapidoth2017}
A. Lapidoth, \emph{A Foundation in Digital Communication}, Cambridge, U.K.: Cambridge Univ. Press, 2009.

\bibitem{Zakai1965}
M. Zakai, ``Band-limited functions and the sampling theorem,'' \emph{Information and Control}, vol. 8, no. 2, pp. 143--158, Apr. 1965.

\bibitem{CT06}
T. Cover and J. A. Thomas, \emph{Elements of Information Theory}, 2nd Ed., Wiley-Interscience, New York, NY, USA, 2006.

\bibitem{RA09}
F. Rusek and J. B. Anderson, ``Constrained capacities for faster-than-Nyquist signaling," \emph{IEEE Trans. Inform. Theory}, vol. 55, no. 2, pp. 764-775, Feb. 2009.

\bibitem{MacKay1999}
D. J. C. MacKay, ``Good error-correcting codes based on very sparse matrices,'' \emph{IEEE Trans. Inf. Theory}, vol. 45, no. 2, pp. 399--431, Mar. 1999.

\end{thebibliography}
\end{document}